\documentclass{acmproc-sp}

\begin{document}
%
% --- Author Metadata here ---
% -- Can be completely blank or contain 'commented' information like this...
%\conferenceinfo{WOODSTOCK}{'97 El Paso, Texas USA} % If you happen to know the conference location etc.
%\setpagenumber{50} % Commented out since no page numbering required.
%\CopyrightYear{2001} % Allows a non-default  copyright year  to be 'entered' - IF NEED BE.
%\crdata{0-12345-67-8/90/01}  % Allows non-default copyright data to be 'entered' - IF NEED BE.
% --- End of author Metadata ---

\title{Quantum Bit Escrow}

%
% You need the command \numberofauthors to handle the "boxing"
% and alignment of the authors under the title, and to add
% a section for authors number 4 through n.
%
% Up to the first three authors are aligned under the title;
% use the \alignauthor commands below to handle those names
% and affiliations. Add names, affiliations, addresses for
% additional authors as the argument to \additionalauthors;
% these will be set for you without further effort on your
% part as the last section in the body of your article BEFORE
% References or any Appendices.

\numberofauthors{3}
%
% You can go ahead and credit authors number 4+ here;
% their names will appear in a section called
% "Additional Authors" just before the Appendices
% (if there are any) or Bibliography (if there
% aren't)

% Put no more than the first THREE authors in the \author command

%% e-mail address with \email.

\author{
        Dorit Aharonov
	\thanks{This research was supported in part by a U.C. president's postdoctoral fellowship and NSF Grant CCR-9800024.} \\
	University of California \\
        Berkeley, CA 94720 \\
        doria@cs.berkeley.edu
\and
        Amnon Ta-Shma 
        \thanks{Supported in part by David Zuckerman's David and Lucile 
Packard Fellowship for Science and Engineering and
NSF NYI Grant No.\ CCR-9457799.}\\
        University of California\\
        Berkeley, CA 94720 \\
        amnon@cs.berkeley.edu 
\and 
        Umesh V. Vazirani 
        \thanks{This research was supported in part by NSF Grant CCR-9800024,
        and a JSEP grant.} \\ 
        University of California \\
        Berkeley, CA 94720 \\
        vazirani@cs.berkeley.edu
\and
        Andrew C. Yao 
        \thanks{This research was supported in part by DARPA and NSF 
          under CCR-9627819, by NSF under CCR-9820855, 
          and by a Visiting Professorship sponsored by 
          the Research Miller Institute at Berkeley.} \\
        Princeton University\\
	Princeton, NJ 08544 \\
        yao@cs.princeton.edu
}

\date{17 February 2000}
\maketitle

%%%%%%%%%%%%%%%%%%%%%%%%%%%%%%%% BEGIN macros %%%%%%%%%%%%%%%%%%%%%%%%%%

\newcommand{\remove}[1]{}

\newtheorem{theorem}{Theorem}   
\newtheorem{remark}{Remark}   
\newtheorem{definition}{Definition}
\newtheorem{lemma}[theorem]{Lemma}
\newtheorem{claim}[theorem]{Claim}
\newtheorem{corollary}[theorem]{Corollary}
\newtheorem{example}{Example}
\newtheorem{attempt}{Attempt}
\newtheorem{protocol}{Protocol}
\newtheorem{cl}[theorem]{Claim}
\newtheorem{fact}{Fact}
\newtheorem{proposition}{Proposition}
\newtheorem{conjecture}{Conjecture}

\newcommand{\prob}{Prob}
\newcommand{\eqdef}{\stackrel{\rm def}{=}} 
\newcommand{\eqq}{\stackrel{\rm ?}{=}} 
\newcommand{\ket}[1]{|{#1}\rangle}
\newcommand{\density}[1]{| {#1} \rangle \langle {#1}| }
\newcommand{\tensor}{ \otimes }
\newcommand{\xor}{ \oplus }
\newcommand{\la}{ \langle }
\newcommand{\ra}{ \rangle }
\newcommand{\co}{{\cal O}}
\newcommand{\ch}{{\cal H}}
\newcommand{\tenhv}{}
\newcommand{\minent}{{H_\infty}}
\newcommand{\set}[1]{\{#1\}}
\newcommand{\trn}[1]{|| {#1} ||_{t}}
\newcommand{\norm}[1]{|| {#1} ||}
\newcommand{\trace}[1]{Trace({#1})}
\newcommand{\ol}[1]{\overline{#1}}

\newcommand{\up}{\uparrow}
\newcommand{\down}{\downarrow}

\newcommand{\inceps}[2]{
\epsfxsize= #2 cm
\centerline{\epsffile{#1.eps}}
}

\def\ccc{{\mathchoice {\setbox0=\hbox{$\displaystyle\rm C$}\hbox{\hbox
to0pt{\kern0.4\wd0\vrule height0.9\ht0\hss}\box0}}
{\setbox0=\hbox{$\textstyle\rm C$}\hbox{\hbox
to0pt{\kern0.4\wd0\vrule height0.9\ht0\hss}\box0}}
{\setbox0=\hbox{$\scriptstyle\rm C$}\hbox{\hbox
to0pt{\kern0.4\wd0\vrule height0.9\ht0\hss}\box0}}
{\setbox0=\hbox{$\scriptscriptstyle\rm C$}\hbox{\hbox
to0pt{\kern0.4\wd0\vrule height0.9\ht0\hss}\box0}}}}

%%%%%%%%%%%%%%%%%%%%%%%%%%%%%%%% end macros %%%%%%%%%%%%%%%%%%%%%%%%%%

\begin{abstract}
Unconditionally secure bit commitment and coin flipping are known 
to be impossible in the classical world. 
Bit commitment is known to be impossible also in the quantum world. 
We introduce a related new primitive - {\em quantum bit escrow}.
In this primitive Alice commits to a bit $b$ to Bob.
The commitment is {\em binding} in the sense that if Alice is asked to 
reveal the bit, Alice can not bias  her commitment without having a 
good probability of being detected cheating.
The commitment is {\em sealing} in the sense that if Bob learns information
about the encoded bit, then if later on he is
asked to prove he was playing honestly,
he is detected cheating with a good probability.
%The difference from Bit commitment is that Bob 
%can learn information about the deposited bit before revealing time, 
%but not without the risk of being caught with certain probability. 
%This is a unique feature of the quantum world. 
Rigorously proving the correctness of quantum cryptographic protocols 
has proved to be a difficult task.
We develop techniques to prove quantitative statements 
about the binding and sealing properties of the quantum bit escrow protocol. 

A related primitive we construct is a quantum biased coin flipping protocol 
where no player can control the game,
i.e., even an all-powerful cheating player 
must lose with some constant probability, 
which stands in sharp contrast to the classical world
where such protocols are impossible.

\end{abstract}

\remove{
% A category with only the three required fields
%A category including the fourth, optional field follows...
}


%\keywords{ACM proceedings, \LaTeX, text tagging} % NOT required for Proceedings

\sloppypar

\section{Introduction}
\label{sec:intro}

We start with an informal definition of a (very) weak variant of bit 
commitment. In this variant there is first a commitment stage in which 
Alice commits a bit $b$ to Bob. Later on there is a reveal stage in which 
Alice reveals the bit and Bob proves he played honestly. The protocol
should be binding in the sense that if Alice changes her mind 
at revealing time then Bob has a good probability of catching her cheating,
and sealing in the sense that if Bob learns information about the committed
bit then Alice has a good probability of catching him cheating. 
Thus, the fundamental (and only) difference between this primitive and 
bit commitment
is that in bit commitment Bob can not learn from the encoding
any information about $b$, while in the weak primitive Bob can learn
a lot of information about the encoded bit, but if he does so Alice
catches him cheating with a good probability.

\begin{definition}(Weak bit commitment)
\label{def:weak}
A weak bit commitment protocol is 
a quantum communication protocol between Alice and Bob 
which consists of two stages, 
the depositing stage and the revealing stage, and a final classical 
declaration stage at which both Alice and Bob 
each declare ``accept'' or ``reject''. 
The following requirements should hold. 
\begin{itemize}
\item
If both Alice and Bob are honest, then at depositing stage
 Alice decides 
on a bit, $b$. She then communicates with Bob, where Alice's 
protocol depends on $b$. At revealing stage Alice 
and Bob communicate, and during this stage 
Alice reveals to Bob the deposited bit $b$.
Both Alice and Bob accept.
\remove{\item 
 qubit does not reveal, in an information theoretic sense,
all the information about the deposited bit $b$.}
\item (Binding)
If Alice tries to change her mind about the value of $b$, then 
there is non zero probability that an honest Bob would 
reject.  
\item (Sealing)
If Bob attempts to learn information about the deposited 
bit $b$, then there is non zero probability that 
an honest Alice would reject. 
\end{itemize}                                                               
\end{definition}

Later on, we will give more formal definitions of 
``Alice changing her mind'' and ``Bob learning information'',
and we will quantify the degree to which a protocol is 
binding or sealing.

Now, consider the following protocol:

\begin{protocol} (Bit Escrow)
\label{pro:escrow}
For an angle $\alpha \in [-\pi,\pi]$
define
$\phi_\alpha = \cos(\alpha) \ket{0}+\sin(\alpha) \ket{1}$.
Let,

\begin{eqnarray*}
\phi_{b,x} & = &
\left\{
\begin{array}{ll}
\phi_{-\theta} & b=0,x=0 \\
\phi_{\theta}  & b=0,x=1 \\
\phi_{{\pi \over 2}-\theta}  & b=1,x=0 \\
\phi_{{\pi \over 2}+\theta}  & b=1,x=1
\end{array}
\right.
\end{eqnarray*}
for some fixed angle $\theta \le {\pi \over 8}$,
say, $\theta={\pi \over 8}$.
See Figure \ref{fig:s0}.

\begin{figure}[t]
\centering
\epsfig{file=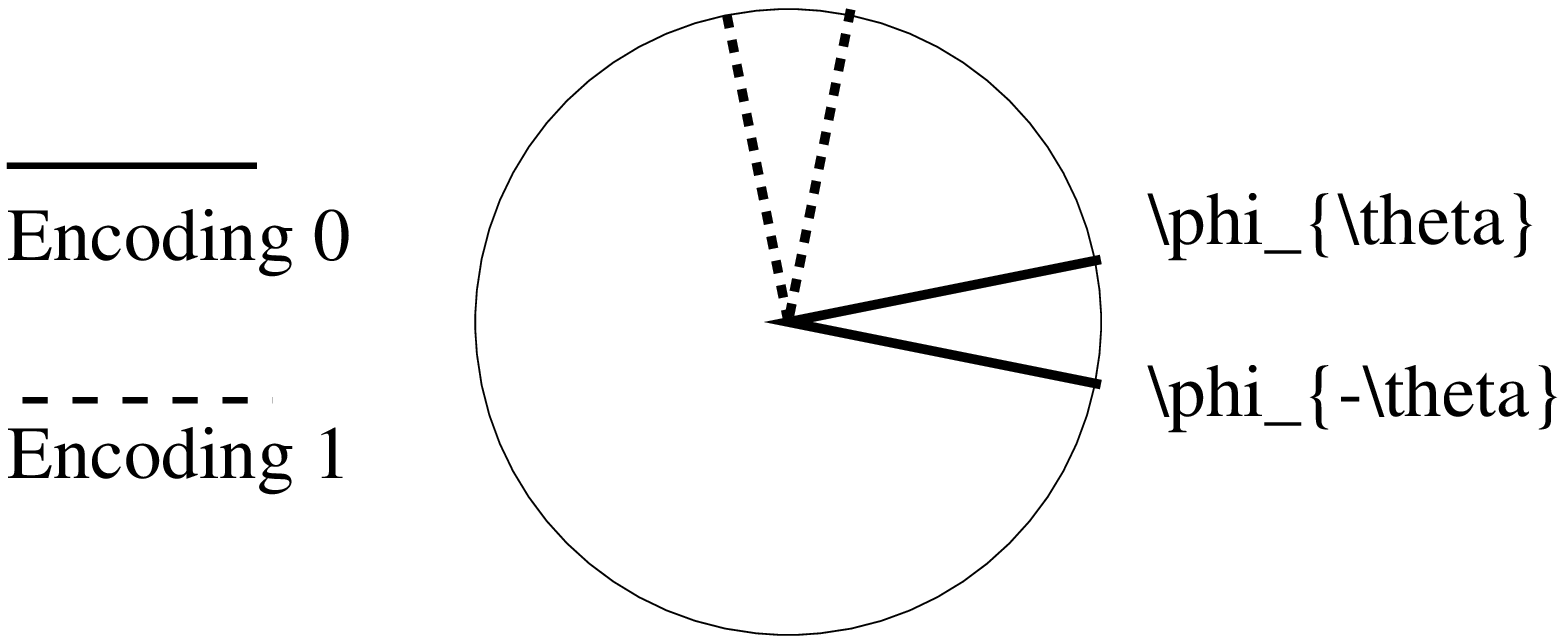,height=1.2in, width=3in}
\caption{\it $\phi_{b,x}$}
\label{fig:s0}
\end{figure}

To deposit bit $b$, Alice picks a random $x \in \{0,1\}$,
and sends $\phi_{b,x}$ to Bob.
Later on, one of the following two challenges is issued:

\begin{itemize}
\item
Either Alice is asked to reveal the deposited bit, 
and then Alice sends the classical bits $b$ and $x$ to Bob
\footnote{
This means that when Bob gets the qubit $q_b$ that is supposed to
carry a classical value for $b$, Bob measures $q_b$ first in
the $\set{\ket{0},\ket{1}}$ basis.
We carry this convention throughout the paper.}.
Bob measures $\phi$ according to the basis
$\set{\phi_{0,x},\phi_{1,x}}$ and verifies that
the result of the measurement is $\phi_{b,x}$.
\item
Or Bob is asked to return the deposited qubit,
he returns a qubit $q$, and Alice measures it in the 
$\set{\phi_{0,x},\phi_{1,x}}$ basis and verifies that it is $\phi_{b,x}$.
\end{itemize}
\end{protocol}

We rigorously define and prove:

\begin{theorem}
\label{thm:escrow}
Protocol \ref{pro:escrow} has  the following properties:

\begin{itemize}
\item 
The deposited qubit does not reveal, in an information theoretic sense,
all the information about the deposited bit $b$.
\item (Binding)
When Bob asks Alice to reveal the classical bit $b$ that she deposited,
if Alice influences the value of b with advantage $\epsilon$ 
then she is detected cheating with probability $\Omega(\epsilon^2)$.
\item (Sealing)
When Alice challenges Bob to return the deposited qubit, then 
if Bob can predict $b$ with advantage $\epsilon$
then he is detected cheating with probability $\Omega(\epsilon^2)$.
\end{itemize}
\end{theorem}

Protocol \ref{pro:escrow} and Theorem \ref{thm:escrow} 
do not achieve the goal set in definition \ref{def:weak}
of weak bit commitment. 
Definition \ref{def:weak} asks for a protocol that is both binding and
sealing, i.e., a commitment s.t. if either player cheats he is detected
cheating with a good probability. Protocol \ref{pro:escrow}
and Theorem \ref{thm:escrow} only give a commitment that is either
binding (if Alice has to reveal) or sealing (if Bob has to return the
qubit), but not simultaneously both.
We therefore call this protocol a {\it bit escrow} protocol. 
The question of achieving simultaneous binding and sealing
 i.e. a weak bit commitment protocol, is left open. 
This question was addressed
in \cite{HK99},
 who independently defined the binding and sealing
properties, and we discuss it in section \ref{sec:related}. 

We describe soon how to use the first two properties
in Theorem \ref{thm:escrow} to get a biased 
coin flipping protocol with a constant bias.

\subsection{Quantum Coin flipping}

Alice and Bob are going through a divorce. They want to decide by a coin 
flip over the phone who is going to keep the car.
The problem is that they do not trust each other any more.

\begin{definition}
(Classical coin flipping)
\cite{B81}
A coin flipping protocol with $\delta$ bias
is one where Alice and Bob communicate 
and finally decide on a value $c \in \{0,1\}$ s.t.
if at least one of the players is honest then 
for any strategy of the dishonest player
$Prob(c=0) \in [{1 \over 2}-\delta,{1 \over 2} + \delta]$.
\end{definition}

Classical coin flipping can be implemented either by a trusted party or
by assuming players with limited computational power and some
cryptographic assumptions. However, if the players have unlimited 
computational power then no coin flipping protocol is possible in a classical
world. This is because any protocol represents a two player game,
and therefore game theory tells us that there is a  player with
an always winning strategy.

By contrast, in the quantum setting coin flipping 
(without computational assumptions)
is not a priori ruled out. 
This is because any attempt by a player to measure 
extra information by deviating from the protocol can 
disturb the quantum state, and 
therefore be detected by the other player.
This leads Lo and Chau\cite{LC98} and later Mayers {\it et. al.}\cite{MSC99}
to consider quantum coin flipping. 
There are several ways to define quantum coin flipping
when cheaters can be detected.
We define:

\begin{definition}
\label{def:non}
(quantum coin flipping)
A quantum coin flipping protocol with bias $\delta$
 is one where Alice and Bob communicate 
and finally each decides on a value $c \in \{0,1,err \}$.
Let $c_A$ ($c_B$) denote Alice's (Bob's) result.
We require: 
\begin{itemize}
\item
If both players are honest then $c_A$ always equals $c_B$,
$\prob(c_A=err)=0$, and $0$ and $1$ have equal probability: 
$Prob(c_A=0)=Prob(c_A=1)={1 \over 2}$.
\item
If one of the players is honest and the other is not,
then for any strategy of the dishonest player,
the honest player's result $c$ satisfies for any $b \in \set{0,1}$:
$$Prob(c=b) \le {1 \over 2}+ \delta$$
\end{itemize}
\end{definition}

Lo and Chau \cite{LC98}
showed that there is no quantum  coin flipping 
protocol with $0$ bias, under a certain restriction
 (``ideal coin flipping''.) 
Mayers et al \cite{MSC99} generalized their proof to the 
general $0$ bias case. 
Lo and Chau leave open the question whether non-exact protocols exist.
Mayers et al \cite{MSC99} suggest a quantum coin flipping protocol
that is based on a biased-coin protocol that is repeated many times.
Mayers et al prove that it works well against some strong, natural 
attacks. However, no general proof is given or
claimed for the coin-flipping protocol
or the biased-coin sub-protocol.

We give a simple protocol for quantum biased coin flipping,
with constant bias.  
It is a modification of protocol \ref{pro:escrow}:

\begin{protocol}(A biased coin flipping protocol)
\label{pro:bias}

\begin{itemize}
\item
Alice picks $b,x \in_R \set{0,1}$ and sends Bob $\phi_{b,x}$.
We set $\theta={\pi \over 8}$.
\item
Bob chooses $b' \in_R \set{0,1}$ and sends it to Alice.
\item
Alice sends Bob $b$ and $x$. Bob checks against the qubit
she sent in the first step. The result of the game is 
$r=err$ if Alice is caught cheating and $r=b \oplus b'$ otherwise.
\end{itemize}
\end{protocol}

Based on the properties of protocol \ref{pro:escrow}  we can prove that 
no player can fully control the game:

\begin{theorem}
\label{thm:bias}
\label{THM:BIAS}
Protocol \ref{pro:bias} has $\delta \le 0.42$ bias.
\end{theorem}

i.e., no player can force his result with probability greater than
$0.92$. We note that while our protocol is resilient against all powerful
malicious quantum players, it requires 
only simple single qubit operations
from the honest player. 
An intriguing question is whether 
quantum coin flipping protocols are possible for 
arbitrarily low biases.

\subsection{Weak Bit Commitment?}
\label{sec:related}
Hardy and Kent \cite{HK99} (see Section \ref{sec:related2})
noticed that Protocol \ref{pro:escrow} can be used to 
give a weak bit commitment protocol 
if Alice and Bob can access a random independent coin flip. 
This is done as follows:  at revealing time 
 Alice first reveals the bit $b$, and then 
they receive a random independent coin flip. 
If the coin is $0$, Bob is challenged to convince Alice that 
he hasn't been cheating, and if the coin flip turns out to be $1$, 
then Alice is challenged. 
This is still correct if the coin flip is biased, as long as both  
probabilities for $0$ and for $1$ are constant.

\begin{figure}[t]
\centering
\epsfig{file=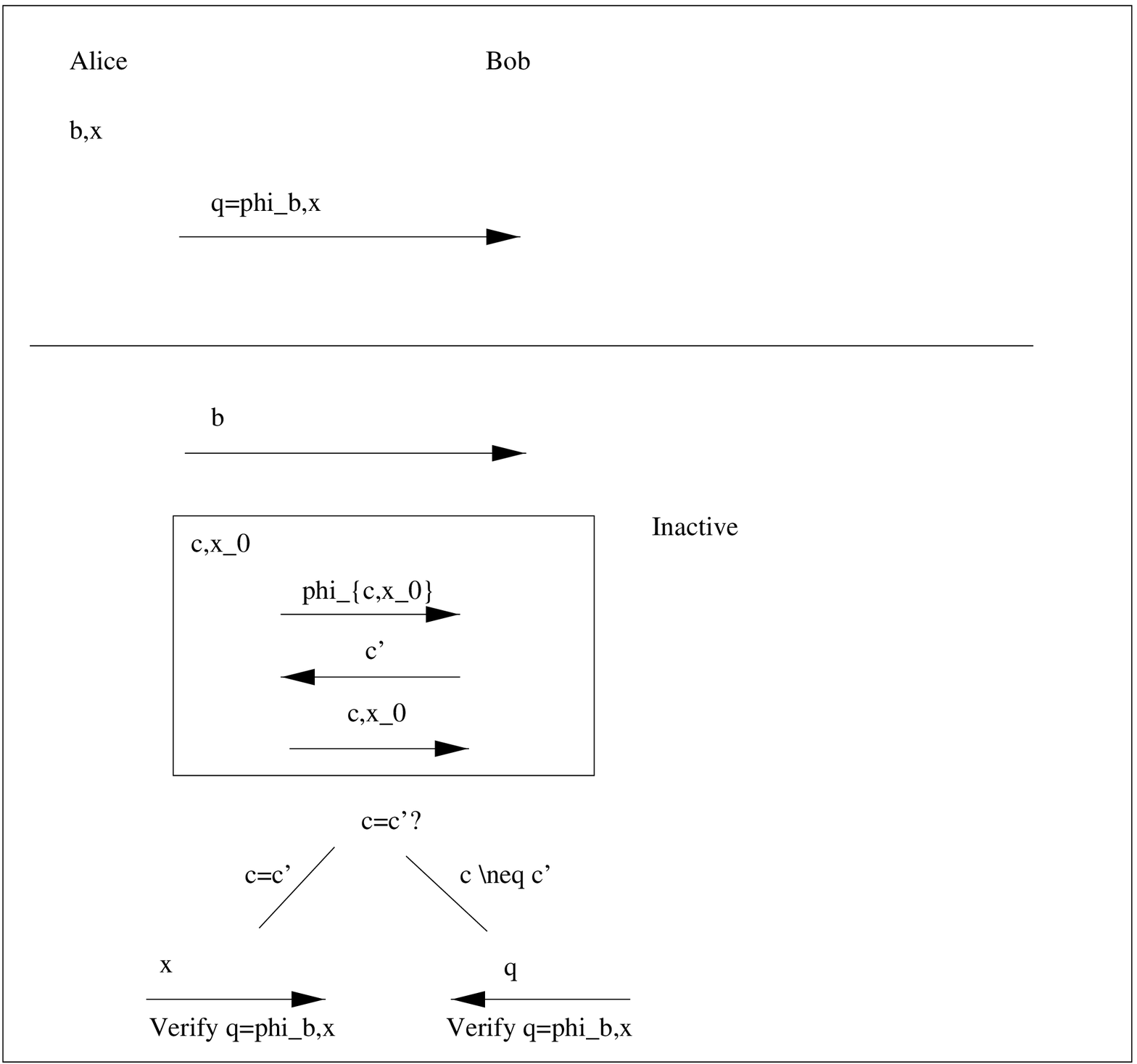,height=4in, width=3in}
\caption{\it Protocol \ref{pro:hk}}
\label{fig:hk}
\end{figure}

Since we already have a biased coin flipping protocol, 
we might consider using this biased coin flipping protocol 
combined with the bit escrow protocol to give a 
weak bit commitment protocol. 
Consider the following protocol (see Figure \ref{fig:hk}):  

\begin{protocol}
\label{pro:hk}
To deposit bit $b$, Alice picks a random $x \in \{0,1\}$,
and sends $\phi=\phi_{b,x}$ to Bob.
To reveal the bit, Alice sends $b$ to Bob. Then a biased-coin
flipping protocol (Protocol \ref{pro:bias}) is played.

\begin{itemize}
\item
If Alice loses she is asked to reveal $x$ and 
Bob measures $\phi$ according to the basis
$\set{\phi_{0,x},\phi_{1,x}}$ and verifies that
the result of the measurement is $\phi_{b,x}$.
\item
If Bob loses he is asked to return the deposited qubit $q$, 
and Alice measures it in the 
$\set{\phi_{0,x},\phi_{1,x}}$ basis and verifies that it is $\phi_{b,x}$.
\end{itemize}
\end{protocol}

It is left as an open question whether this protocol, 
or perhaps a protocol which uses a different coin flipping procedure, 
is actually a weak bit commitment protocol. 
The main difficulty in proving or disproving such a result 
 is the issue of independence between the 
coin flipping protocol and the bit escrow protocol.  
In other words, one has to prove that the 
 cheater cannot use entanglement to correlate the events of 
being detected cheating in the  bit-escrow protocol and 
winning the biased coin flipping protocol, in such a way that 
the cheater is never challenged
when he (or she) has positive probability of being detected.

It is our hope that our techniques could be extended to give weak bit 
commitment
with $\Omega(\epsilon^c)$ binding and sealing for some constant $c$.
Our results also show that Protocol \ref{pro:hk} cannot be more
than $\Omega(\epsilon^2)$ sealing or binding. It might be interesting to 
find a protocol that does better, or prove that such a protocol does not
exist. It seems that a weak bit commitment protocol with 
better than quadratic security parameters can be used repeatedly to give a  
secure coin flipping protocol with unbounded bias.

\subsection{Related Work}
\label{sec:related2}

Some of the work presented here was independently done by 
Hardy and Kent \cite{HK99}. They independently
 defined the binding and sealing
properties and the weak bit commitment primitive (giving it different names).
The protocol they analyze is similar in structure 
to protocol \ref{pro:hk}. 
Hardy and Kent's result asserts that 
a protocol similar to Protocol 
\ref{pro:hk} is simultaneously sealing and binding.
I.e., if Alice (Bob) uses a strategy that gives her (him)
$\epsilon$ advantage, then Alice (Bob) is detected cheating with 
some probability which is strictly greater than $0$  
(they do not analyze the dependence of the detection probability 
on $\epsilon$). 
However, no proof is given regarding the security against 
a cheater who tries to correlate the two parts 
of the protocol to his (or her) advantage.    

\remove{ 
It is our hope that our techniques can be extended to 
 give the following:

\begin{conjecture}
\label{conj:hk}
Protocol \ref{pro:hk}, or a variant of it, 
 is a weak bit commitment protocol with 
 $\Omega(\epsilon^2)$ sealing and binding.
I.e., if Alice (Bob) uses a strategy that gives her (him)
$\epsilon$ advantage, then Alice (Bob) is detected cheating with 
probability $\Omega(\epsilon^2)$.
\end{conjecture}
}

%%%%%%%%%%%%%%%%%%%

\remove{
\begin{definition}
(ideal quantum coin flipping)
\cite{LC98}
An {\em ideal} quantum 
coin flipping protocol is one where Alice and Bob communicate 
and finally decide on a value $c \in \{0,1,err \}$ s.t.
\begin{itemize}
\item
If both players are honest then 'err' never occurs
and $Prob(c=0)=Prob(c=1)={1 \over 2}$.
\item
If one of the players is honest and the other is not,
then whenever the cheater is not caught
the result is $0$ or $1$ with equal probability.
Formally, 
$Prob( c=0~|~c \neq err) = {1 \over 2}$.
\end{itemize}
\end{definition}
}

\remove{The definitions of sealing and binding, and the idea 
of quantum bit commitment which is ``cheat sensitive'', 
were independently discovered by Hardy and Kent\cite{hk}. 
In the original version of this paper, 
 we have proved that Protocol \ref{pro:escrow} is either binding (if Alice is 
asked to reveal the classical data $b$ and $x$) or sealing (if Bob
is asked to return the qubit). Hardy and Kent do better 
in that they prove that protocol \ref{pro:hk}
 is simultaneously binding and revealing.
On the other hand, Hardy and Kent prove only that if Alice (or Bob) cheats,
then there is some positive detection probability, which might be }

\section{Preliminaries}
\label{sec:pre}

\noindent
{\bf The model}.
Let $\set{ e_1,\ldots,e_{2^n} }$ be an orthonormal basis for $\ccc^n$,
and let $\ket{i} = \ket{i_1,\ldots,i_n}$ be the vector $e_i$.
A pure state over $n$ qubits is a vector $v \in \ccc^{2^n}$ of norm $1$.
Any pure state $\ket{v}$ can be expressed as 
$ \ket{v} = \Sigma_i a_i \ket{i}$, with $\Sigma_i |a_i|^2 = 1$.
A mixed state is a classical distribution over pure states, 
$\{p_i,\phi_i\}$,  where
$0 \le p_i \le 1$, $\Sigma_i p_i=1$ and $\phi_i$ is a pure state,
and the interpretation we give it is that the system is with probability
$p_i$ in the pure state $\phi_i$.
A quantum system is, in general, in a mixed state.
The system Alice builds in the first stage of Protocol \ref{pro:bias}
is in a mixed state that is 
with probability ${1 \over 4}$ in some pure state $\phi_{b,x}$.

A quantum system can undergo two basic operations:
unitary evolution and measurement.

\begin{description}
\item[Unitary evolution]:
If a unitary transformation $U : \ccc^{2^n} \mapsto \ccc^{2^n}$
is applied to a pure state $\phi$,
then the new state of the system is the pure state
$U \phi$.
If $U$ is applied to the mixture
$\{p_i,\phi_i\}$ then the new state of the system is
the mixture $\{p_i, U \phi_i \}$. 
The interpretation we give it is that with probability $p_i$
the system was in the pure state $\phi_i$ hence it is now in the
pure state $U \phi_i$.

\item[Orthogonal Measurements]:
An orthogonal  measurement is a decomposition of the system into 
orthogonal subspaces.
More formally, 
suppose the system is in a super position $\phi \in \ccc^{2^n}$.
Suppose $\ch_1,\ldots,\ch_k$ are orthogonal subspaces,
and $\ccc^{2^n}= ~\ch_1 \oplus \ldots \oplus \ch_k$.
A measurement of $\phi$ according to the decomposition
$\ch_1,\ldots,\ch_k$,
will get result $i$ (or $\ch_i$) with probability
$q_i=|\Pi_{\ch_i} |\phi\ra|^2 $ where $\Pi_{\ch_i}$ 
is the projection on subspace  $\ch_i$, and then the state will collapse
to ${1 \over \sqrt{q_i} } \Pi_{\ch_i} |\phi \ra$.
In other words, $\phi$ falls into the subspace $\ch_i$ with probability 
which is the length of the projection squared, and the new
vector is the normalized projected vector.
An orthogonal measurement can be represented using an Hermitian 
matrix $M$ whose  eigenspaces are the subspaces $\ch_i$.
%and the 
%corresponding eigenvalues are the classical outcomes of the measurement, 
%i.e. the labels, $i$, of the subspace to which the state is projected.      
A measurement of a mixture is the mixture of the measurements of the 
pure states.
\end{description}

Given a system $\rho$ on $\ccc^{2^n}$,
one can use an ancilla, say $\ket{0,\ldots,0} \in \ccc^{2^m}$,
apply a unitary transformation 
$U: \ccc^{2^n} \otimes \ccc^{2^m} \mapsto \ccc^{2^n} \otimes \ccc^{2^m}$,
and then an orthogonal measurement on $\ccc^{2^n} \otimes \ccc^{2^m}$.
It turns out that this is the most general measurement possible.
There are several equivalent ways to formulate this so called
'generalized measurement', and we refer the interested reader to 
\cite{P98}.

{~}

\noindent
{\bf The Density Matrix.}
The density matrix of a pure state $\ket{\phi}$ is the matrix
$\density{\phi}$, where $\la \phi |= ((\phi)^t)^*$
is the conjugate transpose of $\phi$.
For example, the density matrix of $\phi_{0,0}$ is
\begin{align*}
\density{\phi_{-\theta}} & = & 
%\left(
%\begin{array}{cc}
%\cos(\theta) \\
%-\sin(\theta) 
%\end{array}
%\right)
%\left(
%\begin{array}{cc}
%\cos(\theta) & -\sin(\theta) 
%\end{array}
%\right)
%~=~
\left(
\begin{array}{cc}
\cos^2(\theta) & -\cos(\theta) \sin(\theta) \\
-\cos(\theta) \sin(\theta) & \sin^2(\theta)
\end{array}
\right)
\end{align*}
The density matrix of a mixed state 
$\{p_i,\phi_i\}$ 
is $\Sigma_i p_i \density{\phi_i}$.
All density matrices are Hermitian, positive semi-definite
and have trace $1$.
If a unitary matrix $U$ operates on the system, 
it transforms the density matrix $\rho$ to 
$U\rho U^\dagger$. 
A measurement $M$ operating on a system whose density matrix 
is $\rho$ results in an expected outcome $Trace(M\rho)$.  

{~}

\noindent{\bf Distinguishing Between Density Matrices.}
Given a quantum system $\rho$ and a generalized measurement
$\co$ on it, let $\rho^{\co}$ denote the classical distribution 
on the possible results that we get by measuring $\rho$ according to 
$\co$. i.e., it is some classical distribution $p_1,\ldots,p_k$
where we get result $i$ with probability $p_i$.
Given two different mixed states, we can ask how well one can 
distinguish between the two mixtures.
We need a measure for the distance between two classical distributions and 
we choose the $l_1$ norm:

\begin{definition}
Let $p_1,\ldots,p_k$ and $q_1,\ldots,q_k$ be two probability
distributions over $\{1,\ldots,k\}$.
Then 
$|p-q|_1 = \Sigma_i |p_i-q_i|$.
\end{definition}

A fundamental theorem about distinguishing density matrices\cite{AKN98}
 tells us:

\begin{theorem}
\cite{AKN98}
\label{thm:akn}
Let $\rho_1,\rho_2$ be two density matrices on the same space $\ch$.
Then for any generalized measurement $\co$
$$|\rho_1^\co - \rho_2^\co|_1 \le \trace{\sqrt{A^\dagger A}}$$
where $A=\rho_1-\rho_2$.
Furthermore, the bound is tight, and the
orthogonal measurement $\co$ that projects a state on the 
eigenvectors of $\rho_1-\rho_2$ achieves this bound.
\end{theorem}

Theorem \ref{thm:akn} shows that the 
density matrix captures all the accessible information that a quantum
state contains. If two different mixtures have the same density matrix (which
is quite possible) then physically they are two different systems,
but practically (and from a computational point of view)
they are indistinguishable.

The quantity $\trace{\sqrt{A^\dagger A}}$ is of independent interest.
If we define 
$\trn{A} = \trace{\sqrt{A^\dagger A}}$
then $\trn{\cdot}$ defines a norm, and has some additional properties
such as $\trn{A \tensor B} = \trn{A} \cdot \trn{B}$,
$\trn{A}=1$ for any density matrix $A$ and
$\trn{AB},\trn{BA} \le \trn{A} \cdot \trn{B}$.
If $\phi_1,\phi_2$ are two pure states, and
$\rho_i$ is the reduced density matrix of $\phi_i$, then
$\trn{\rho_0-\rho_1} = 2 \sqrt{1-|\la \phi_1 | \phi_2 \ra|^2}$.
See \cite{AKN98} for more details.

\remove{
\begin{example}\cite{AKN98}
\label{ex:akn}
Suppose $\phi_1,\phi_2$ are two pure states. 
Let $\rho_i$ be the reduced density matrix of $\phi_i$.
$\trn{\rho_0-\rho_1} = 2 \sqrt{1-|\la \phi_1 | \phi_2 \ra|^2}$.

Without loss of generality, let 
$\phi_1 = 
\left( 
\begin{array}{r} 
1 \\ 
0 
\end{array} 
\right) 
$
and
let 
$\phi_2 = 
\left( 
\begin{array}{r} 
\cos(\theta) \\ 
\sin(\theta)
\end{array} 
\right) 
$
Then the optimal
measurement to distinguish $\phi_1$ and $\phi_2$ first rotates 
$\phi_1$ and $\phi_2$ by $\pi/4 - \theta/2$ and then measures in
the $0-1$ basis. The $L_1$ distance between the two probability 
distributions thus obtained is exactly 
$2 \sqrt{1- \cos^2 \theta}$.
\end{example}
}

\remove{
\begin{example}
\label{ex:x}
Let us take Protocol \ref{pro:escrow}.
Assume Alice picks $b \in \set{0,1}$ s.t.
$\Pr(b=0)={1 \over 2}+\delta$ ($\delta \ge 0$)
and $x \in_R \set{0,1}$ and computes $\phi_{b,x}$.
We compute the reduced density matrix $\rho_{x=0}$ of the event $x=0$.
I.e., with probability ${1 \over 2}+\delta$ 
we have $b=0$ and the system is in $\phi_{0,0}$,
and with probability ${1 \over 2}-\delta$ 
we have $b=1$ and the system is in $\phi_{1,0}$.
Therefore, 
\begin{eqnarray*}
\rho_{x=0} & = &
%({1 \over 2}+\delta)
%\left(
%\begin{array}{cc}
%cos^2(\theta) & -sin(\theta) cos(\theta) \\
%-sin(\theta)cos(\theta) & sin^2(\theta)
%\end{array}
%\right)+
%({1 \over 2}-\delta)
%\left(
%\begin{array}{cc}
%sin^2(\theta) & sin(\theta)cos(\theta) \\
%sin(\theta)cos(\theta) &  cos^2(\theta)
%\end{array}
%\right) \\
%& = &
{1 \over 2}
\left(
\begin{array}{cc}
1 & 0 \\
0 & 1
\end{array}
\right)
+ 
\delta
\left(
\begin{array}{cc}
cos(2\theta) & -sin(2\theta) \\
-sin(2\theta) & -cos(2\theta)
\end{array}
\right)
\end{eqnarray*}
Similarly,
The reduced density matrix $\rho_{x=1}$ of the event $x=1$ is
$
{1 \over 2}
\left(
\begin{array}{cc}
1 & 0 \\
0 & 1
\end{array}
\right)
+ 
\delta
\left(
\begin{array}{cc}
cos(2\theta) & sin(2\theta) \\
sin(2\theta) & -cos(2\theta)
\end{array}
\right)$.
It follows that 
$A=\rho_0-\rho_1 = 
\delta
\left(
\begin{array}{cc}
0 & -2sin(2\theta) \\
-2sin(2\theta) & 0
\end{array}
\right)$.
By Theorem \ref{thm:akn}, for any measurement $\co$,
$|\rho_{x=1}^\co - \rho_{x=2}^\co|_1 \le Trace (\sqrt{A^\dagger A})
\le 4 \delta \sin(2\theta)$.
It then follows that $\prob(d=x) \le {1 \over 2}+ \delta \sin(2\theta)$. 
In particular, when $\delta=0$, 
$\prob(d=x)={1 \over 2}$ so the state contains no information at all
about $x$.
\end{example}
}

\remove{
\begin{example}
\label{ex:b}
We continue with the previous example.
Let $\rho_{b=0}$ be the density matrix in the case $b=0$,
and $\rho_{b=1}$ in the case $b=1$. Then 
$\rho_{b=0}=
\left(
\begin{array}{cc}
cos^2(\theta) & 0 \\
0 & sin^2(\theta)
\end{array}
\right)$
and
$\rho_{b=1}=
\left(
\begin{array}{cc}
sin^2(\theta) & 0 \\
0 & cos^2(\theta)
\end{array}
\right)$.
Hence, 
$\rho_{b=0}-\rho_{b=1} =
\left(
\begin{array}{cc}
cos(2\theta) & 0 \\
0 & -cos(2\theta)
\end{array}
\right)$.
It follows that:
$\Pr(c=b) \le {1 \over 2}+{\cos(2 \theta) \over 2}$,
or equivalently,
$\Pr(c \ne b) \ge {1 - \cos(2 \theta) \over 2} = sin^2(\theta)$.
\end{example}
}

{~}

\noindent
{\bf Locality.}
We now turn to the local view of a subsystem.
Suppose we are in a mixed state  $\rho$ over $k+m$
qubits, where Alice holds the first $k$ qubits $A$
and Bob holds the last $m$ qubits $B$.
Assume that Alice applies a generalized measurement $\co$ on her qubits $A$.
This induces a new density matrix $\rho_B^{\co}$ on $B$.
E.g., if Alice and Bob were in the super position
$\phi={1 \over \sqrt{2}} (\ket{00}+\ket{11})$ over two qubits
and Alice measured the second qubit
according to the basis $\{ \ket{0},\ket{1}\}$, then 
Bob is with probability ${1 \over 2}$ in the super position
$\ket{0}$ and with probability ${1 \over 2}$ in $\ket{1}$,
hence 
$\rho_B^{\co} =
\left(
\begin{array}{cc}
{1 \over 2} & 0 \\
0 & {1 \over 2}
\end{array}
\right)
$.
A fundamental fact from physics,
which can also be proven rigorously, tells us that 
in fact $\rho_B^{\co}$ does not depend on $\co$, but only 
on the original matrix $\rho$. 
We thus denote it by $\rho|_B$, and call it the
  density matrix $\rho$ reduced onto the subsystem $B$.
Alternatively, we say that the rest of the system is {\it traced out}. 
The physical interpretation of the above result 
is that a player is guaranteed locality,
i.e., a player Bob who holds a subsystem $B$ knows that 
the results he gets from measurements he applies on $B$
do not depend on the way the system outside $B$ evolves.
It is also some kind of commitment.
If Alice sends Bob $k$ qubits that have reduced density matrix
$\rho_B$, then whatever Alice later does can not change this 
reduced density matrix. 

{~}

\noindent
{\bf Purification.}
A density matrix on a Hilbert space $A$ can always be viewed as a 
reduced density matrix of a pure state on a larger Hilbert space, 
a process which is called ``purification''. 
A pure state $|\phi\ra_{A,B}$ is a purification of the density matrix
 $\rho_A$ if the reduced density matrix of $|\phi\ra\la \phi|_{A,B}$
to the Hilbert space $A$ is $\rho$. 
The most straight forward way to purify a density matrix 
  $\rho=\sum_i w_i |\phi_i\ra\la \phi_i|$ 
is by the state $|\phi\ra=\sum_i \sqrt{w_i} |i\ra\otimes |\phi_i\ra$.

{~}

\noindent
{\bf Fidelity.}

 The fidelity is a way to measure distances between density matrices, 
which is an alternative to the trace metric. 
Given two density matrices $\rho_0,\rho_1$ 
 on the same Hilbert space 
$A$ the fidelity is defined \cite{J94} to be: 
\begin{equation}
f(\rho_0,\rho_1)=\sup|\la\phi_0|\phi_1\ra|^2
\end{equation}
where the supremum is taken over all purifications $|\phi_0\ra$ of $\rho_0$
and $|\phi_1\ra$ of $\rho_1$ to the same dimensional Hilbert space. 
We note here a few important properties which can easily be proven:
\begin{enumerate}                                             
\item $0\le f(\rho_0,\rho_1)\le 1$
\item $f(\rho_0,\rho_1)=1 \Longleftrightarrow \rho_0=\rho_1$
\item For $\rho_0$ which is a pure state, i.e. 
 $\rho_0=|\phi_0\ra\la\phi_0|$, we have \[ f(\rho_0,\rho_1)=\la \phi_0|\rho_1|\phi_0\ra.\]
\end{enumerate}
Note that the fidelity increases as the distance between 
 two density matrices decreases. 
It is also not too difficult to see that the supremum is always achieved, 
i.e. we can replace the supremum by a maximum; 
See \cite{J94} for more details. 

{~}

\noindent
{\bf Entanglement.}
Suppose Alice holds a register $A$, Bob holds $B$, and the 
system is in a pure state $\psi_{AB}$. If we look at Bob's system alone
then we might see a mixed state, and
as we said before, Alice can not change the reduced density matrix of Bob
by local operations on her side. 
On the other hand Alice might gain 
different aspects of knowledge on the actual result that Bob gets.

\begin{example}
$\psi_{AB}={1 \over \sqrt{2}}(\ket{00}+\ket{11})$.
If Alice measures in the $\set{\ket{0},\ket{1}}$ basis, then 
Bob's system is with probability half
in the state $\ket{0}$, and with probability half in the state $\ket{1}$,
and the register $A$ reflects the result Bob gets, i.e., Alice knows
whether Bob gets a zero or a one.
Now, $\psi_{AB}$ can also be represented as
${1 \over \sqrt{2}}(\ket{+,+}+\ket{-,-})$
where $|+\ra={1 \over \sqrt{2}}(\ket{0}+\ket{1})$
and $|-\ra={1 \over \sqrt{2}}(\ket{0}-\ket{1})$.
Alice can measure the register $A$ in the $\set{|+\ra,|-\ra}$ basis.
Now Bob's system is with probability ${1 \over 2}$
in the state $\ket{+}$, and with probability half in the state 
$\ket{-}$,
and the register $A$ reflects the result Bob gets, i.e., Alice 
knows whether Bob gets $\ket{+}$ or $\ket{-}$.
Notice that Bob's reduced density matrix is the same in both cases.
\end{example}

An important Theorem by Mayers \cite{M97} 
and independently Lo and  Chau \cite{LC98} states:

\begin{theorem}
\label{thm:lo}
Suppose the reduced density matrix of $B$ is the same in 
$\phi_{AB}$ and $\psi_{AB}$. Then
Alice can move from $\phi_{AB}$ to $\psi_{AB}$
by applying a {\em local} transformation on her side.
\end{theorem}

I.e., even though Alice can not change Bob's reduced density matrix,
she can determine how to ``open'' the mixture, and do so in a way 
that gives her full knowledge of Bob's result.

\section{The Binding Property}

In Protocol \ref{pro:escrow} Alice sends a qubit to Bob (we call it
a ``deposit'' step) and later on she tells Bob how to ``open''
the qubit (the ``reveal'' step) 
which also determines the value that is supposed to be in the qubit.
Such a protocol is worthless unless the deposit step is ``binding''
Alice to a pre-determined value.
We first define the binding property in a general way.
We then analyze how binding
Protocol \ref{pro:escrow} is. Suppose we have a two step protocol:

\begin{description}
\item[Deposit]:
Alice prepares a super-position 
$\psi_{AB}$ with two quantum registers $A$ and $B$.
Alice sends the second register $B$ to Bob.

\item[Reveal]:
Alice and Bob communicate.
Bob follows the protocol and Alice is arbitrary.
If Alice wants to create a bias towards $0$ she uses one strategy,
and if she wants a bias towards $1$ she uses a different strategy.
Bob decides on a  result $r_B \in \{0,1,err\}$.
\end{description}

Let us denote by $p_0$ the probability that Alice claims the result is $0$
in the zero strategy, 
by $p_1$ the probability that Alice claims the result is $1$
in the zero strategy, 
and by $p_{err}$ the probability
that Bob decides the answer is $r_B=err$ when Alice uses
the zero strategy.
We similarly define 
$q_0,q_1,q_{err}$ for the one strategy.

\begin{definition} ($(\epsilon,\gamma)$ binding)
A protocol is $(\epsilon,\gamma)$ binding,
if whenever Bob is honest, for any strategy Alice uses, 
if $p_{err},q_{err} \le \epsilon$
then $|p_0-q_0|,|p_1-q_1| \le \gamma$.
%
%We also say that Alice can not $\epsilon$-change her commitment
%with failure probability less than $p$.
\end{definition}

\subsection{Protocol \ref{pro:escrow} is quadratically binding}

\begin{theorem}
\label{thm:binding}
Protocol \ref{pro:escrow} is 
$(\epsilon,\gamma= {2\sqrt{\epsilon} \over \cos(2 \theta)})$ binding.
\end{theorem}

\begin{proof}
(of Theorem \ref{thm:binding}).
At deposit time Alice sends Bob one qubit $B$,
which might be entangled with the qubits $A$ that Alice holds.
Let us denote the reduced density matrix of $B$ by $\rho$.
At revealing time, Alice may choose whether she wants 
to bias the result towards $0$,
in which case she applies the generalized measurement $M_0$,
or towards $1$ in which case she applies $M_1$.
The measurements $M_0$ and $M_1$ do not change the reduced density matrix 
$\rho$ of Bob, but rather give different ways to realize $\rho$ as a mixture
of pure-states, and give Alice information about the value that Bob
actually gets to see in this mixture.

Now, we even go further and give Alice complete freedom to choose the way
she realizes the reduced density matrix $\rho$ of Bob as a mixture,
and we give her the knowledge of Bob's value for free.
Let us say that when Alice applies $M_0$, 
the reduced density matrix $\rho$ 
is realized as the mixture $\set{p_i, \phi_i}$, 
and when Alice applies $M_1$ 
the reduced density matrix $\rho$ 
is realized as the mixture $\set{p_i', \phi_i'}$. 

Now, let us focus on the zero strategy.
Say Alice realizes $\rho$ as $\set{p_i,\phi_i}$.
When the $i$'th event  happens,
Alice's strategy tells her to send some two qubits $q_b,q_x$ to Bob,
that are supposed to hold classical $0,1$ values for $b$ and $x$. 
Bob then measures $q_b$ and $q_x$ in the $\set{\ket{0},\ket{1}}$ basis.
Now, if one of $q_b,q_x$ is not a classical bit, 
then Alice can measure it herself in the $\set{\ket{0},\ket{1}}$ basis,
and get a mixture over classical bits.
Furthermore, we can push all the probabilistic decisions into the 
mixture $\set{p_i,\phi_i}$.
Thus, w.l.o.g, we can assume Alice's answers $q_b$ and $q_x$
are classical bits that are determined by the event $i$.
Let us denote by $u_i$ the vector $\phi_{b_i,x_i}$ where
$b_i,x_i$ are Alice's answers when event $i$ occurs.
W.l.o.g we may assume $u_i \in \set{\phi_{b,x}}$, otherwise we know
Bob immediately rejects.

The probability Bob discovers that Alice is cheating is then
$1-|\la \phi_i | u_i \ra|^2$ and the overall probability
Bob detects Alice is cheating is

\begin{eqnarray*}
p_{err} & = & \Sigma_i p_i (1-|\la \phi_i | u_i \ra|^2) 
\end{eqnarray*}

Let us define the density matrix 
$\rho_0 = \Sigma_i p_i | u_i \ra \la u_i|$.

\begin{claim}
$\trn{\rho - \rho_0}  \le 2 \sqrt{p_{err}}$.  
\end{claim}

\begin{proof}
$\trn{|\phi_i \ra \la \phi_i| - |u_i \ra \la u_i|} ~=~
2 \sqrt{1-|\la \phi_i | u_i \ra|^2}$.
Therefore

\begin{eqnarray*}
\trn{\rho - \rho_0}  & = & 
\trn{\Sigma_i p_i |\phi_i \ra \la \phi_i| - 
      \Sigma_i p_i |u_i \ra \la u_i|~} \\
 & \le & \Sigma_i p_i \trn{|\phi_i \ra \la \phi_i| - 
                          |u_i \ra \la u_i|~} \\
 & =   & 2 \Sigma_i p_i \sqrt{1-|\la \phi_i | u_i \ra|^2}
\end{eqnarray*}

Now, by Cauchy-Schwartz inequality,

\begin{eqnarray*}
\Sigma_i p_i \sqrt{1-|\la \phi_i | u_i \ra|^2} &=&
\Sigma_i \sqrt{p_i} \sqrt{p_i (1-|\la \phi_i | u_i \ra|^2)} \\
 &\le&
\sqrt{\Sigma_i p_i} \sqrt{\Sigma_i p_i (1-|\la \phi_i | u_i \ra|^2)} \\
&=& \sqrt{p_{err}}
\end{eqnarray*}

and the claim follows.
\end{proof}

Similarly, if Alice tries to bias the result towards $1$,
$B$ ends up in the mixture $\{p_i',\phi_i'\}$,
and when $\phi_i'$ occurs Alice sends $b',x'$
to Bob that correspond to a vector $u_i' \in \set{\phi_{b,x}}$.
We define $\rho_1$ to be the reduced density matrix 
$\rho_1=\Sigma_i p_i' |u_i' \ra \la u_i'|$.
As before, 
$ \trn{ \rho - \rho_1 }   \le 2 \sqrt{q_{err}}$.
Hence,
$ \trn{ \rho_0 - \rho_1 }   \le 2(\sqrt{p_{err}}+\sqrt{q_{err}})$.

To conclude the proof, we establish the following claim:

\begin{claim}
Let $\rho_0$ and $\rho_1$ be density matrices corresponding to mixtures 
over $\set{\phi_{b,x}}$. Let 
$p_0$ be the probability of $\phi_{0,0}$ or $\phi_{0,1}$ 
in the first mixture,
and $p_1 = 1 - p_0$ be the probability of $\phi_{1,0}$ or $\phi_{1,1}$.
Similarly let $q_0$ and $q_1$ be the corresponding quantities for
the second mixture. Then 
$\trn{\rho_0-\rho_1} \geq 2 \cdot |p_0 - q_0| \cos 2\theta$.
\end{claim}

\begin{proof}

We show that we can distinguish the mixtures with probability at least 
$|p_0 - q_0| \cos 2\theta$ 
when we measure them according to the basis $\set{\ket{0},\ket{1}}$.
If we do the measurement on a qubit whose state is
the reduced density matrix $\rho_0$ we get the
$\ket{0}$ answer with probability $p_0 \cos^2(\theta) + p_1 \sin^2(\theta)$,
while if we do the measurement on a qubit whose state is
the reduced density matrix $\rho_1$ we get the
$\ket{0}$ answer with probability 
$q_0 \cos^2(\theta) + q_1 \sin^2(\theta)$. 
The difference is
$| p_0 \cos^2(\theta) + p_1 \sin^2(\theta) -
 (q_0 \cos^2(\theta) + q_1 \sin^2(\theta))| 
= |p_0-q_0| (\cos^2(\theta) - \sin^2(\theta)) $,
where we used $p_1-q_1=(1-p_0)-(1-q_0)=q_0-p_0$.  
Altogether we get
$
\trn{\rho_0 -\rho_1} ~\ge~
2 \cdot |p_0-q_0| (\cos^2(\theta) - \sin^2(\theta)) 
$
as desired.
\end{proof}

Putting it together:
$$2 \cdot \cos(2 \theta) \cdot |p_0-q_0| ~\le~ \trn{\rho_1-\rho_1} 
\le 2(\sqrt{p_{err}}+\sqrt{q_{err}}) \le 4 \sqrt{\epsilon}$$
I.e., 
$|p_0-q_0| \le {2\sqrt{\epsilon} \over \cos(2 \theta)}$.

\end{proof}

\subsection{A Quadratic Strategy for Alice}

We now show that Alice has a quadratic strategy
for Protocol \ref{pro:escrow}, and thus Theorem \ref{thm:binding} is
essentially tight. In fact, we show the quadratic bound for a more general
family of protocols.
Let $\rho_0, \rho_1$ be two density matrices of the same 
dimension, $\rho_0$ can be realized as the mixture 
$\set{p^0_i,\ket{\alpha^0_i}}$, and $\rho_1$ as
$\set{p^1_i,\ket{\alpha^1_i}}$.
To encode $b$, honest Alice picks $\ket{\alpha^b_i}$ with probability
$p_i$ and sends it to Bob. At revealing time Alice sends $b$ and $i$ to Bob,
and Bob tests whether Alice is cheating by projecting his state on 
$\ket{\alpha^b_i}$. 

\begin{theorem}
\label{thm:quadraticalice}
Let $f$ be the fidelity $f(\rho_0,\rho_1)$.
For any $0\le \alpha \le \pi/4$ there exists a strategy for Alice
with advantage $\sqrt{f} sin(2\alpha)/2$ and probability of 
detection at most ${(1-f) sin^2(\alpha) \over 2}$. 
\end{theorem}

On first reading of the next proof the reader might want to 
check the proof in the simpler case where $\rho_0$ and $\rho_1$
represent pure states, i.e.,
$\rho_b = | \psi_b \ra\la \psi_b|$.

\begin{proof}
We first represent the strategy of a honest Alice in quantum language.  
Consider two maximally parallel purifications 
 $\ket{\psi_0}$ and $\ket{\psi_1}$ of $\rho_0$ and $\rho_1$,
where $\rho_0$ and $\rho_1$ are density matrices of the register $B$, 
and the purifications are states on a larger Hilbert space 
 $A\otimes B$. 
By \cite{J94}, $|\la \psi_0|\psi_1\ra|^2=f(\rho_0,\rho_1)$. 
At preparation time, Alice prepares the state 
\begin{eqnarray*}
\label{alicestate}
\ket{\beta} & = & \frac{1}{\sqrt{2}}(\ket{0,\psi_0}+\ket{1,\psi_1})
\end{eqnarray*}
on $A\otimes B$ and one extra qubit $C$. 
Alice then sends the register $B$ to Bob. 
At revealing time, Alice measures the qubit $C$
in the $\ket{0},\ket{1}$ basis, to get a bit $b$. 
The state of registers $A,B$ is now $\ket{\psi_b}$. 
Alice then applies a unitary transformation $U_b$ on register $A$, 
which  rotates her state $\ket{\psi_b}$ to the state 
\begin{eqnarray*}
\ket{\psi'_b} &=& \sum_j \sqrt{p^b_j} |j\ra_A|\alpha^b_j\ra_B
\end{eqnarray*}
This is possible by Theorem \ref{thm:lo}. 
After applying $U_b$, Alice measures register $A$ in the computational 
basis and sends Bob the bit $b$ and the outcome of the second measurement, 
$j$.  
This strategy is similar to the honest strategy, except 
for that Alice does not know what bit and state 
is sent until revealing time. 

We can also assume w.l.o.g. that the maximally parallel purifications satisfy 
that $\la\psi_0|\psi_1\ra$
is real and positive. This can be assumed since otherwise we could multiply 
$|\psi_0\ra$ by an overall phase without changing the reduced density 
matrix and the absolute value of the inner product. 

To cheat, Alice creates the encoding $\ket{\beta}_{CAB}$ 
and sends register $B$ to 
Bob. Alice's one strategy is also as described above.
The zero strategy, on the other hand,
is a slight modification of the honest strategy. 
At revealing time, Alice measures the control 
qubit $C$ in the $\set{\ket{\phi_\alpha},\ket{\phi_\alpha^\perp}}$
basis, where 
\begin{eqnarray}
\ket{\phi_\alpha}&=&c|0\ra+s|1\ra,\\\nonumber 
\ket{\phi_\alpha^\perp}&=& -s|0\ra+c|1\ra,
\end{eqnarray}
and $s=\sin(\alpha)$, $c=\cos(\alpha)$.
 If the outcome is a projection on $\ket{\phi_\alpha}$
Alice sends $b=0$ and proceeds according to the $b=0$ honest protocol, 
i.e. applies $U_0$ to register $A$, 
measures in the computational basis and sends the result to Bob. 
If the outcome is a projection on  $\ket{\phi_\alpha^\perp}$, 
Alice proceeds according to the $b=1$ honest protocol.  
Let us now compute Alice's advantage and Alice's probability 
of getting caught cheating. 

We can express $\ket{\beta}$ as:
\begin{eqnarray*}
\ket{\beta} & = & 
{1 \over \sqrt{2}} 
 (c \ket{\phi_\alpha,\psi_0} -s \ket{\phi_\alpha^\perp,\psi_0}) + \\
& & {1 \over \sqrt{2}} 
 (s \ket{\phi_\alpha,\psi_1} +c \ket{\phi_\alpha^\perp,\psi_1}). 
\end{eqnarray*}
 Hence, the probability
Alice sends $b=0$ in the zero strategy is 
${1 \over 2} |c \psi_0+s \psi_1|^2 = 
{1 \over 2}(c^2+s^2+2cs \la \psi_0 | \psi_1 \ra) = 
{1 \over 2}(1+2cs\sqrt{f})$.
We conclude:

\begin{claim}
Alice's advantage is ${\sqrt{f} \sin(2\alpha) \over 2}$.
\end{claim}

We now prove that the detection probability is at most $(1-f)s^2$.
The  state of $A\otimes B$  
conditioned that the first measurement yields $\ket{\phi_\alpha}$
can be written as
 $\frac{1}{\sqrt{Pr(b=0)}}{1 \over \sqrt{2}}(c \ket{\psi_0} + s \ket{\psi_1})$
where $Pr(b=0)$ is the probability Alice sends $b=0$ in the zero 
strategy. The above state can be written as 
$$
\frac{1}{\sqrt{Pr(b=0)}}{1 \over \sqrt{2}}(c+\sqrt{f}s) \ket{\psi_0}+
\sqrt{1-f}s \ket{\psi_0^\perp}
$$
The rest of the protocol involves Alice's rotation of the state by $U_0$, 
then Alice's measurement of the register $A$ and Bob's 
measurement of the register $B$.
The entire process can be treated as a generalized measurement on 
this state, where this measurement is a projection onto one of two subspaces, 
the ``cheating Alice'' and the ``Honest Alice'' subspaces. 
We know that $\ket{\psi_0}$ lies entirely in the honest Alice subspace, 
and thus the probability that Alice is caught, conditioned that 
$C$ was projected on $\phi_\alpha$, is at most
$\frac{1}{Pr(b=0)} {1 \over 2}(1-f)s^2$.

In the same way, when we condition on a projection on $\phi_\alpha^\perp$, 
Alice's state can be written as  
$\frac{1}{\sqrt{Pr(b=1)}}{1 \over \sqrt{2}}
((c-\sqrt{f}s) \ket{\psi_1}-\sqrt{1-f}s \ket{\psi_1^\perp})$.
which gives a probability of detection which is at most 
$\frac{1}{Pr(b=1)}{1 \over 2} (1-f)s^2$.
Adding the conditional probabilities together we get that 
the detection probability is at most
${(1-f)s^2 \over 2}$.
\end{proof}

%%%%%%%%%%%%%%%%%%%% OLD %%%%%%%%%%%%%%%%%%%%%%%%%%%%%%%%%%%%
%%%%%%%%%%%%%%%%%%%%%%%%%%%%%%%%%%%%%%%%%%%%%%%%%%%%%%%%%%%%%

\remove{
We prove will prove that the protocol \ref{pro:escrow}
is quadratically binding, i.e. that Alice cannot get advantage which 
is larger than square root of her probability to be caught. 
To illustrate the techniques of the proof of this theorem, as well as the 
difficulties that motivate them, let us first look at an example
for a
protocol in which Alice has positive advantage 
with zero probability of being caught. 
Such a bad scenario does not occur in our protocol.

\subsection{Examples}

First, we consider what happens 
in Protocol \ref{pro:escrow} 
if Alice follows a naive strategy
of trying to open a $0$ as a $1$. Say that she sent the state 
$\phi_{0,1}=\cos \theta \ket{0} + \sin \theta \ket{1}$ to Bob, 
so Alice can always open the bit as a zero.
Now suppose she tries
to open it as a $1$ by telling Bob that she actually sent the 
state 
$\phi_{1,0}=\cos (\pi/2 - \theta) \ket{0} + \sin(\pi/2 - \theta) \ket{1}$.
Now Bob makes his measurement on the
escrow qubit to verify her claim.
Bob is convinced to a one
(i.e. Alice successfully cheats) with probability
$\cos^2 (\pi/2 - 2\theta)$, 
and will catch the error with a constant probability
$\cos^2 \theta$.  So, Alice can not perfectly cheat using this strategy.

The real issue is whether or not Alice
can follow a more subtle strategy: one where the qubit she
deposits with Bob is entangled with qubits that she holds on to,
and such that if she measures her qubits in a certain basis,
she gets exactly the same result that Bob gets in that same basis. 
If she could arrange this, she could still get Bob to open a $0$-deposit
as a $1$ with probability $\cos^2 (\pi/2 - 2\theta)$, but now 
would not have the corresponding probability $\cos^2 \theta$
of being caught cheating. Indeed, this issue is quite subtle,
and below we illustrate a small variant of our escrow protocol in
which Alice can indeed follow such a strategy.

We now describe a specific protocol.
Let the sets 
$T_0=\{ v_1=\phi_{-\theta},v_2=\phi_0,v_3=\phi_{\theta} \}$
and
$T_1=\{ v_4=\phi_{{\pi \over 2} -\theta},v_5=\phi_{{\pi \over 2}},
        v_6=\phi_{{\pi \over 2}+\theta} \}$
be as in Figure \ref{fig:t0}.
At deposit time Alice is supposed to pick a vector $v$ of her choice
from $T_0 \cup T_1$ and send it to Bob.
At reveal time, Alice has to send the angle $\alpha$ s.t. 
$v=\phi_{\alpha}$. Bob then measures his qubit according to the 
basis $\set{\phi_{\alpha},\phi_{\alpha}^\bot}$.
If the result of the measurement is $\phi_{\alpha}^\bot$ 
the protocol ends with an $err$ result.
Otherwise, if $\phi_{\alpha} \in T_0$ the result is $0$,
if $\phi_{\alpha} \in T_1$ the result is $1$,
and if $\phi_{\alpha} \not \in T_0 \cup T_1$ the result is $err$.

\begin{figure}[t]
\begin{center}
%\epsfxsize=6.4in
%\hspace{0in}
%\epsfbox{2to1.eps}
\inceps{t0}{8}
\end{center}
\caption{\it Encoding one bit using $T_0$ and $T_1$}
\label{fig:t0}
\end{figure}

\remove{
We now check several strategies for a cheating Alice.

\begin{attempt}
\mbox{ }\\
{\bf Deposit.}
Alice prepares the super-position
$\ket{0,\phi_{0}}+\ket{1,\phi_{\pi \over 2}}$
and sends the second qubit to Bob. \\
{\bf Reveal.}
Alice measures her qubit, if it is $0$ 
she sends Bob  $b=0,\alpha=0$ 
otherwise
she sends Bob  $b=1,\alpha={\pi \over 2}$.  
\end{attempt}

In this example $p_{err}=q_{err}=0$ and $p_0=p_1=q_0=q_1={1 \over 2}$.
This strategy just amounts to a delayed coin flip.
At revealing time Alice declares $0$ with
probability half and $1$ with probability half,
and she might as well have tossed the coin $b$ at deposit time.

\begin{attempt}
\mbox{ }\\
{\bf Deposit.}
Alice sends $\phi_{\pi/4}={1 \over {\sqrt{2}}} (\ket{0}+\ket{1})$ to Bob.\\
{\bf Reveal.}
If Alice wants to convince Bob to $0$ she sends him 
$b=0,\alpha=\theta$, 
and if she wants to convince Bob to $1$ she sends him 
$b=1,\alpha={\pi \over 2} - \theta$.
\end{attempt}

Now $p_0=1,q_0=0$ and $p_1=0,q_1=1$
but $p_{err}=q_{err}=1-\cos^2(\pi/8) \approx 0.14$.
Thus, in this example Alice gets a substantial bias towards her value.
The price is, however, that there is a constant probability $p_{err}$
she is caught cheating.
}

In our example
we build two different mixtures of $\{v_1,\ldots,v_6\}$
that yield the same density matrix.
It can be checked that the mixture $\{p_i,v_i\}$ with
$p=(0,p,0,{1 \over 4},{1 \over 2}-p,{1 \over 4})$
has the same density matrix as that of the
mixture $\{p_i',v_i\}$ with
$w'=({1 \over 4},0,{1 \over 4},0,{1 \over 2},0)$,
when 
$p={\cos^2(\theta)-\sin^2(\theta) \over 2} = {\cos(2\theta) \over 2}$.
Notice, however, that while in the first mixture the 
probability of a vector from the zero region is 
$p_1+p_2+p_3 = p = {\sqrt{2} \over 4} \approx 0.353$,
in the second mixture the probability of a vector from the
zero region is $p_1'+p_2'+p_3'= {1 \over 2}$.

\remove{
$\left(
\begin{array}{cc}
p+{\sin^2(\theta) \over 2} & 0 \\
0 & {1 \over 2}-p + {\cos^2(\theta) \over 2} 
\end{array}
\right)
$.
$\left(
\begin{array}{cc}
{\cos^2(\theta) \over 2} & 0 \\
0 & {1 \over 2} + {\sin^2(\theta) \over 2} 
\end{array}
\right)
$.
}

\begin{example}
\mbox{ }\\
{\bf Deposit.}
Let  $\{e_1,\ldots,e_8\}$ be an orthonormal basis for $\ccc^8$.
Alice prepares the super-position
$\psi^0_{AB}={1 \over 2} \ket{e_1,v_1} + {1 \over 2} \ket{e_3,v_3} + 
 {1 \over \sqrt{2}} \ket{e_5,v_5}$   
and sends the qubit $B$ to Bob.\\
{\bf Reveal.}
If Alice wants to convince Bob to $0$ she measures her qubits $A$ in the
$E=\{e_1,\ldots,e_8\}$ basis.
If she gets $e_1$ she sends $b=0,\alpha=-\theta$, 
if she gets $e_3$ she sends $b=0,\alpha=\theta$ and
if she gets $e_5$ she sends $b=1,\alpha=\pi/2$.
If Alice wants to convince Bob to $1$ she does a {\em local} unitary
transformation that transforms the super-position $\psi^0_{AB}$
to the super-position 
$\psi^1_{AB} = 
           {1 \over {\sqrt{p}}} \ket{f_2,v_2} + {1 \over 2} \ket{f_4,v_4}
        + {1 \over \sqrt{1/2-p}} \ket{f_5,v_5} + {1 \over 2} \ket{f_6,v_6}$.
Such a {\em local} transformation exists by Theorem \ref{thm:lo}. 
Alice then measures in the $F$ basis. 
If she gets $f_2$ she sends $b=0,\alpha=0$, 
if she gets $f_4$ she sends $b=1,\alpha=\pi/2-\theta$,
if she gets $f_5$ she sends $b=1,\alpha=\pi/2$ and
if she gets $f_6$ she sends $b=1,\alpha=\pi/2+\theta$.
\end{example}
 
It can be easily seen that Alice is never caught cheating, thus
$p_{err}=q_{err}=0$.
Also, 
$p_0={1 \over 2}$ and $q_1=1-p \approx 0.64$.
The last example demonstrates an extremely bad scenario:
Alice can cheat and get a significant bias, without ever risking being
caught. We prove that this does not happen with protocol \ref{pro:escrow}. 
}

%%%%%%%%%%%%%%%%%%%

\remove{
W.l.o.g., we can assume that  $\la\psi_0|\psi_1\ra$ is real and positive, 
since otherwise we can multiply $\ket{\psi_0}$ by an  overall 
phase which cannot be measured by Bob.   
We also simplify the strategy of a honest Alice in this case.  
At preparation time, Alice prepares the state 
\begin{eqnarray*}
\label{statealice}
\ket{v} &=&
\frac{1}{\sqrt{2}}( \ket{0,\psi_0} + \ket{1,\psi_1} )
\end{eqnarray*}
and sends the second register $B$ to Bob. 
At revealing time, Alice measures her qubit $A$, and sends the result to Bob. 

To gain a quadratic advantage, Alice slightly modifies this strategy. 
At committing time, she behaves as if she is honest by preparing 
the state $\ket{v}$ and sending the second register to Bob. 
Her advantage is gained at revealing time. 
In her one strategy she acts honestly and measures her 
qubit $A$ according to the $\set{\ket{0},\ket{1}}$ basis,
while in her zero strategy she measures $A$ 
in the basis $\set{\phi_{\alpha},\phi_{\alpha}^\perp}$
which is rotated from the computational basis by an angle $\alpha$.
If the state is projected on $\ket{\phi_{\alpha}}$
Alice sends Bob $b=0$,otherwise $b=1$. 

Expressing $\ket{v}$ in the rotated basis we get:
\begin{eqnarray*}
\ket{v} & = & 
{1 \over \sqrt{2}} (
 \ket{c \phi_\alpha -s \phi_\alpha^\perp,\psi_0} + \\
& &
 \ket{
  s \phi_\alpha + c \phi_\alpha^\perp,
  f \psi_0+\sqrt{1-f^2}\psi_0^\perp})
\end{eqnarray*}
where $c=\cos(\alpha),s=\sin(\alpha),f=|\la\psi_0|\psi_1\ra|$.
Rearranging,
\begin{eqnarray*}
\ket{v} & = & 
{1 \over \sqrt{2}} ( 
(c+f s)  \ket{\phi_\alpha,\psi_0}  +\\
& &~~~~~~~~~         
\sqrt{1-f^2} s  \ket{\phi_\alpha,\psi_0^\perp}  +\\
& &~~~~~~~~~         
              (-s+fc)  \ket{\phi_\alpha^\perp,\psi_0}  +      \\
& &~~~~~~~~~         
              \sqrt{1-f^2} c    \ket{\phi_\alpha^\perp,\psi_0^\perp})
\end{eqnarray*}

We see that in her zero strategy Alice sends $0$ with probability
${(c+fs)^2+(1-f^2)s^2 \over 2}=
  {1 \over 2}+csf$,
so Alice's advantage is $csf={f \sin(2\alpha) \over 2}$. 

If $\ket{v}$ collapses to $\phi_\alpha$ then Alice claims
$b=0$ and is caught cheating with probability $(1-f^2)s^2$.
If, on the other hand $\ket{v}$ collapses to $\phi_\alpha^\perp$
then the new super-position is
$(-s+fc) \ket{\psi_0}  + 
\sqrt{1-f^2} c    \ket{\psi_0^\perp}$ normalized.
The probability Alice is caught cheating is 
$[ (-s+fc) |\la \psi_0       | \psi_1^\perp \ra|  + 
  \sqrt{1-f^2} c   |\la \psi_0^\perp | \psi_1^\perp \ra ]^2
~=~
 [ (-s+fc) \sqrt{1-f^2} - \sqrt{1-f^2} c  f ]^2
~=~
s^2 (1-f^2)$.
All together the detection probability is $(1-f^2)s^2$.

We now return to the general case and the proof of Theorem 
\ref{thm:quadraticalice}.
}
\section{The Sealing Property}

\begin{definition} ($(\epsilon,p)$ sealing)
A bit escrow protocol is $(\epsilon,p)$ sealing,
if whenever Alice is honest
and deposits a bit $b$ s.t. $\prob(b=0)={1 \over 2}$,
for any strategy Bob uses
and a value $c$ Bob learns, 
it holds that either
\begin{itemize}
\item
$\Pr_{b \in_R \{0,1\},protocol}~(c=b) \le {1 \over 2}+\epsilon$, or
\item
$\Pr_{b \in_R \{0,1\},protocol}~(r_A=err) \ge p$
\end{itemize}
The probability is taken over $b$ taken uniformly
from $\{0,1\}$ and the protocol.
\end{definition}
We show here that protocol \ref{pro:escrow} 
is quadratically sealing. This means that whatever Bob does, 
he will always be detected cheating with probability which is at least 
the square of his advantage. 
Later, we show that this is tight. 
 
\subsection{Protocol \ref{pro:escrow} is Quadratically Sealing}
\begin{theorem}
\label{thm:sealing}
Protocol \ref{pro:escrow} is 
$(\epsilon=O({\sqrt{p} \over \sin(2\theta)}),p)$ sealing.
\end{theorem}

\begin{proof}
We first describe a general scenario.
Alice is honest and sends $\ket{\phi_{b,x}}_A$ to Bob.
Bob has an ancilla $\ket{0}_C$. Bob applies some unitary transformation
$U$ acting on the registers $A$ and $C$. 
Let us denote 
\begin{eqnarray*}
\ket{\alpha_{b,x}} &=& U( \ket{\phi_{b,x},0}_{AC}) 
\end{eqnarray*}
Bob then sends register $A$
to Alice, and keeps register $C$ to himself.
We want to show that if $C$ contains much information about $b$ then
Alice detects Bob cheating with a good probability.

We can express $\alpha_{b,x}$ as a superposition,   
\begin{eqnarray}
\label{eqn:alpha}
\ket{\alpha_{b,x}} &=& 
\ket{\phi_{b,x},w_{b,x}} +\ket{\phi_{\neg b,x},w'_{b,x}} 
\end{eqnarray}
where we have used the basis $\ket{\phi_{b,x}}$, $\ket{\phi_{\neg b,x}}$,
for $A$.
In this representation, the probability $p$ Bob is caught cheating is:

\begin{eqnarray}
\label{eqn:detection}
p &=& \frac{1}{4} \sum_{b,x} \norm{w'_{b,x}}^2 
\end{eqnarray}
which in particular implies that $\norm{w'_{b,x}} \le 2\sqrt{p}$.

We now want to express Bob's advantage.
Let $\rho_0$ ($\rho_1$) be the reduced density matrix of the register 
$B$ conditioned on the event that $b=0$ ($b=1$). Then, 

\begin{equation}
\label{eqn:rhos}
\rho_b = \sum_{x} Pr(x)
( |w_{b,x}\ra\la w_{b,x}|+ |w'_{b,x}\ra\la w'_{b,x}| )
\end{equation}

Bob's advantage is at most the trace distance between 
$\rho_0$ and $\rho_1$, and we want to bound it from above. 
Triangle inequality gives: 
$\trn{\rho_0-\rho_1} \le
{1 \over 2}
(~ \trn{ |w_{0,0}\ra\la w_{0,0}|- |w_{1,1}\ra\la w_{1,1}|} 
+ \trn{ |w_{0,1}\ra\la w_{0,1}| -|w_{1,0}\ra\la w_{1,0}|} 
+ \sum_{b,x} \trn{w'_{b,x}\ra\la w'_{b,x}|}~)
$.

As the trace norm of two pure states $a$ and $b$ is 
$2\sqrt{1-|\la a|b \ra|^2}$, and
using Equation \ref{eqn:detection}, we get:
\begin{eqnarray*}
\trn{ \rho_0-\rho_1 } 
&\le& \sqrt{1-|\la w_{0,0}| w_{1,1}\ra|^2} + \\
&   & \sqrt{1-|\la w_{0,1}| w_{1,0}\ra|^2} + 2p
\end{eqnarray*}

We now claim;

\begin{lemma}
\label{lem:small}
$|\la w_{0,0}| w_{1,1}\ra| ~,~ |\la w_{0,1}| w_{1,0}\ra| 
~\ge~ 1-O(ctg^2(2\theta)+4)p$.
\end{lemma}

Thus, altogether,
$\trn{\rho_0 - \rho_1} \le O(ctg(2\theta) \sqrt{p})$
which completes the proof.
\end{proof}

We now turn to the proof of Lemma \ref{lem:small}.
\begin{proof} (of Lemma).

We will prove that all the unprimed $w$ vectors lie 
in one bunch of small width, using  the unitarity of $U$. 
The unitarity of $U$ implies that
$\la \phi_{b,x} | \phi_{b',x'} \ra  =
 \la \alpha_{b,x} | \alpha_{b',x'} \ra $.
We can express $\alpha_{b,x}$ as in Equation \ref{eqn:alpha}.
We get:
\begin{eqnarray*}
\la \phi_{b,x}      | \phi_{b',x'} \ra & = & 
\la \phi_{b,x}      | \phi_{b',x'} \ra  \la     w_{b,x} | w_{b',x'} \ra + \\ 
& & 
\la \phi_{b,x}  | \phi_{\neg b',x'} \ra \la w_{b,x} | w'_{b',x'} \ra + \\
& & 
\la \phi_{\neg b,x} | \phi_{b',x'} \ra  \la     w'_{b,x} | w_{b',x'} \ra + \\
& &
\la \phi_{\neg b,x} | \phi_{\neg b',x'} \ra \la w'_{b,x} | w'_{b',x'} \ra  
\end{eqnarray*}
 
Substituting the values $b,x,b',x'$ for actual values, and noticing that
$|\la w'_{b,x} | w'_{b',x'} \ra| \le 4p$, 
we in particular get the 
following equations:

\begin{eqnarray}
\label{eq:wbx}
\la w_{b,x} | w_{b,x} \ra &=_{4p}& 1 \\
\label{eq:w0010}
\la w_{0,0} | w_{1,0}' \ra + \la w_{0,0}' | w_{1,0}\ra &=& 0  \\
\label{eq:w0111}
\la w_{0,1} | w_{1,1}' \ra + \la w_{0,1}' | w_{1,1}\ra &=& 0  
\end{eqnarray}

\begin{eqnarray}
\label{eq:w1011}
\la w_{1,0} | w_{1,1}' \ra - \la w_{1,0}' | w_{1,1} \ra =_{4cp/s} 
{c \over s} (1- \la w_{1,0} | w_{1,1} \ra ) & & \\
\label{eq:w0110}
\la w_{0,1} | w_{1,0}' \ra + \la w_{0,1}' | w_{1,0} \ra =_{4sp/c} 
{s \over c} (1- \la w_{0,1} | w_{1,0} \ra ) & & \\
\label{eq:w0011}
-\la w_{0,0} | w_{1,1}' \ra - \la w_{0,0}' | w_{1,1} \ra =_{4sp/c} 
{s \over c} (1- \la w_{0,0} | w_{1,1} \ra ) & & \\
\label{eq:w0001}
\la w_{0,0}' | w_{0,1} \ra - \la w_{0,0} | w_{0,1}' \ra =_{4cp/s} 
{c \over s} (1- \la w_{0,0} | w_{0,1} \ra) & &
\end{eqnarray}
where $c=\cos(2\theta)$,
$s=\sin(2\theta)$ and we write $x=_qy$ if $|x-y| \le q$.
A partial result can already be derived from what we have so far. 
 By equation \ref{eqn:detection}, we note that the length of the 
primed $w$ vectors is 
 at most  $2\sqrt{p}$. Inserting this to equations 
\ref{eq:w0110} and \ref{eq:w0011}, we get that 
$|\la w_{0,0}| w_{1,1}\ra|$ and similarly 
$|\la w_{0,1}| w_{1,0}\ra|$ are close to $1$ up to terms of order
 $\sqrt{p}$. This is a weaker than the result which we want to achieve
in lemma \ref{lem:small}, 
which is closeness to $1$ up to order $p$ terms. 
If we stop here, the closeness of the unprimed $w$ vectors up 
to order $\sqrt{p}$ implies that Bob's information is at most 
of the order of $^4\sqrt{p}$.  
Note, however, that so far all we have used is unitarity, 
and we have not used the particular properties of the set of vectors 
we use in the protocol.  
In the rest of the proof, we will use the symmetry in protocol 
\ref{pro:escrow} to improve on this partial result, 
and to show that Bob's information is at most of the order of $\sqrt{p}$. 
Basically, the symmetry which we will use is the fact that 
the vectors in the protocol can be paired into orthogonal vectors.

We proceed as follows. The idea is to express equations
\ref{eq:w1011}-\ref{eq:w0001} as inequalities involving only 
the distances between two $w$ vectors, $\norm{w_{b,x}-w_{b',x'}}$
 and then to solve the set of the four inequalities to give an upper bound
on the pairwise distances. This will imply a bound on
the inner products,  $\la  w_{b,x}| w_{b',x'} \ra$, by the following 
connection: 
\begin{claim}\label{real}
$1-Re(\la  w_{b,x}| w_{b',x'} \ra) \ge {\norm{w_{b,x}-w_{b',x'}}^2 \over 2}$.
\end{claim}
where $Re(z)$ denotes the real part of the complex number $z$.  
\begin{proof} 
$\norm{w_{b,x}-w_{b',x'}}^2 = 
\la  w_{b,x}-w_{b',x'} | w_{b,x}-w_{b',x'} \ra \le
2-2Re(\la  w_{b,x}| w_{b',x'} \ra)$.
\end{proof}

We denote:

\begin{eqnarray*}
a &=& \norm{w_{0,0}-w_{0,1}} \\
b &=& \norm{w_{0,0}-w_{1,1}} \\
c &=& \norm{w_{0,1}-w_{1,0}} \\
d &=& \norm{w_{1,0}-w_{1,1}} 
\end{eqnarray*}

Let $LHS$ ($RHS$) be the sum of the left (right) hand side of the last
four equations. 

\begin{claim}
\label{cl:RHS}
$Re(RHS) \ge {c \over 2s} (a^2+d^2) + {s \over 2c} (b^2+c^2)$.
\end{claim}

\begin{proof}
\begin{eqnarray*}
Re(RHS) &=& 
{c \over s}(2-Re(\la w_{0,0} | w_{0,1} \ra) - Re(\la w_{1,0} | w_{1,1} \ra) ) \\
& + & 
{s \over c}(2-Re(\la w_{0,0} | w_{1,1} \ra) - Re(\la w_{0,1} | w_{1,0} \ra) )
\end{eqnarray*}
and now we can apply claim \ref{real}.  
\end{proof}

Expressing the left hand side of the equations in terms 
of $a,b,c$ and $d$ might look a bit more 
complicated, and this is where we invoke the symmetric 
properties of the protocol, namely equations \ref{eq:w0010} 
and \ref{eq:w0111}. 

\begin{claim}
\label{cl:LHS}
$Re(LHS) \le 4 \sqrt{p}(a+b+c+d)$
\end{claim}

\begin{proof}
We first look at the LHS 
of Equation \ref{eq:w0011} + Equation \ref{eq:w0001}.
By adding 
%\la w_{0,1} | w_{1,1}' \ra - \la w_{0,1} | w_{1,1}' \ra
%+ \la w'_{0,1} | w_{1,1} \ra - \la w'_{0,1} | w_{1,1} \ra 
$\la w_{0,1} | w_{1,1}' \ra + \la w'_{0,1} | w_{1,1} \ra = 0$
(due to Equation \ref{eq:w0111})
and by using the fact that
$Re(\la \alpha|\beta\ra)=Re(\la \beta|\alpha\ra)$
we get that the LHS of these two equations contributes
$Re(\la w_{0,0}' | w_{0,1} - w_{1,1}) +
 Re(\la w_{0,1}' | w_{1,1} - w_{0,0}) +
 Re(\la w_{1,1}' | w_{0,1} - w_{0,0}) \le 
2\sqrt{p} (c + d) + 2\sqrt{p} b + 2\sqrt{p} a$.

Similarly, the LHS  of Equation \ref{eq:w1011} +Equation \ref{eq:w0110}
is
$ Re(\la w_{1,0}' | w_{0,1} - w_{1,1}) +
  Re(\la w_{0,1}' | w_{1,0} - w_{1,1}) +
  Re(\la w_{1,1}' | w_{1,0} - w_{0,1})
\le 2\sqrt{p} (a+b +d+c) $.

Altogether, $Re(LHS) \le 4\sqrt{p}(a+b+c+d)$.
\end{proof}

Combining Claims \ref{cl:LHS} and \ref{cl:RHS} with our knowledge that
$ Re(RHS) \le Re(LHS)+ {8cp \over s}+{8sp \over c}$
we get:
\begin{eqnarray*}
& & {c \over 2s} (a^2+d^2) + {s \over 2c} (b^2+c^2)  \le  \\
& & 4 \sqrt{p} (a+b+c+d) + {8cp \over s} +{8sp \over c} 
\end{eqnarray*}

We want to show that $a,b,c,d$ are all of the order of $\sqrt{p}$. 
Define $\Delta=a+b+c+d$. For $0\le \theta\le \frac{\pi}{8}$, 
$ctg(2\theta)\ge tg(2\theta)$. Since all terms in the left hand side 
are positive, we have for each of $a,b,c,d$ an upper bound
in terms of $\Delta$:  
\begin{eqnarray*}
a^2,b^2,c^2,d^2 &\le & {8 \Delta \sqrt{p} \over {s/c}} + 
                       16p(1+({c \over s})^2)
\end{eqnarray*}

Thus, 
$\Delta = a + b+ c + d \le 
4 \sqrt{ {8 \Delta \sqrt{p} c \over s} + { 16p \over s^2}}$.

Solving the quadratic equation
\begin{eqnarray*}
\Delta^2-{2^7 \sqrt{p} c \over s} \Delta - {2^8 p \over s^2} & \le & 0
\end{eqnarray*}
for $\Delta$ we get
\begin{eqnarray*}
\Delta & \le &  132 \cdot \sqrt{p} \cdot ctg(2\theta)
\end{eqnarray*}

Finally,
\begin{eqnarray*}
|\la w_{0,0} |w_{1,1} \ra | & \ge &  |Re( \la w_{0,0} | w_{1,1} \ra)| \\
& = & { \norm{w_{0,0}}^2+\norm{w_{1,1}}^2-b^2 \over 2} \\
& \ge & {2-b^2-8p \over 2} \\
& \ge & 1-(2^{15} ctg^2(2\theta)+4)p
\end{eqnarray*}
where the third inequality is true due to equation \ref{eq:wbx}. 
Similarly, we have the same lower bound for $|\la w_{0,1} |w_{1,0} \ra |$, 
which implies lemma \ref{lem:small}. 
\end{proof}

\remove{
\remove{
We would like to say that all the big $W$ vectors lie in a bunch, 
which is of width of the order of $\sqrt{p}$. 
However, there are 
two inner products missing: $\la W_{0,0}| W_{1,0}\ra$
and $\la W_{0,1}| W_{1,1}\ra$. Moreover, trying to estimate 
these inner products using the above method we are bound to fail 
 since $\la \phi_{0,0}| \phi_{1,0}\ra=
\la \phi_{0,1}| \phi_{1,1}\ra=0$. Fortunately, the property of being
 close up to terms of order $\sqrt{p}$ is transitive (if used finitely many 
times):
\begin{equation}
\la W_{0,0}|W_{1,0}\ra =  \la W_{0,0}|W_{1,1}\ra+ 
\la W_{0,0}|(|W_{1,0}\ra-|W_{1,1}\ra)
\end{equation}
and using the fact that if  $|\la W_{1,0}|W_{1,1}\ra|\ge 1-x $ then 
$\||W_{1,0}\ra-|W_{1,1}\ra\|\le \sqrt{2x}$, we have:
\begin{eqnarray}
&&|\la W_{0,0}| W_{1,0}\ra|\ge1-\\\nonumber
ctg(2\theta)(|\la w_{1,1}| W_{0,0}\ra|+
|\la W_{1,1}| w_{0,0}\ra|)-4p-\\\nonumber
&&\sqrt{2(tg(2\theta)(|\la w_{1,0}| W_{1,1}\ra|+
|\la W_{1,0}| w_{1,1}\ra|)+4p)}. 
\end{eqnarray}

We will now use equation to prove a stronger result, i.e. 
that $\la W_{1,1}| W_{0,0}\ra$ is equal to $1$ up to terms which are of 
order $p$.  
For that, we first write down some of the unitarity restrictions. 
From the restrictions $\phi_{0,0}| \phi_{1,0}\ra
=\la\alpha_{0,0}| \alpha_{1,0}\ra$ and 
$\la\phi_{0,1}| \phi_{1,1}\ra=\alpha_{0,1}| \alpha_{1,1}\ra$
we get 
\begin{eqnarray}
0&=&\la w_{0,0}|W_{1,0}\ra+
\la W_{0,0}|w_{1,0}\ra \\\nonumber
0&=&\la w_{0,1}|W_{1,1}\ra+
\la W_{0,1}|w_{1,1}\ra,  
\end{eqnarray}
or 
\begin{eqnarray}
-\la W_{1,0}|w_{0,0}\ra^*&=&
\la W_{0,0}|w_{1,0}\ra \\\nonumber
-\la W_{1,1}|w_{0,1}\ra^*&=&
\la W_{0,1}|w_{1,1}\ra,  
\end{eqnarray}
and from 
$\la\phi_{0,0}| \phi_{0,1}\ra=\la\alpha_{0,0}| \alpha_{0,1}\ra$ and 
$\la\phi_{0,0}| \phi_{1,1}\ra=\la\alpha_{0,0}| \alpha_{1,1}\ra$ we get
\begin{eqnarray}
cos(2\theta)&=&cos(2\theta)(\la W_{0,0}|W_{0,1}\ra+\la w_{0,0}|w_{0,1}\ra) 
+\\\nonumber&&
sin(2\theta)(\la w_{0,0}|W_{0,1}\ra-
\la W_{0,0}|w_{0,1}\ra)\\\nonumber
-sin(2\theta)&=&sin(2\theta)(-\la W_{0,0}|W_{1,1}\ra+\la w_{0,0}|w_{1,1}\ra)
+\\\nonumber
&&cos(2\theta)(\la w_{0,0}|W_{1,1}\ra+
\la W_{0,0}|w_{1,1}\ra)\\\nonumber
\end{eqnarray}
If we divide the first equation by $sin(2\theta)$ and the second by 
$-cos(2\theta)$ and add the two equations, 
we get: 
\begin{eqnarray}
&&ctg(2\theta)+tg(2\theta)=\\\nonumber
&&ctg(2\theta)\la W_{0,0}|W_{0,1}\ra+ tg(2\theta)\la W_{0,0}|W_{1,1}\ra
-\\\nonumber
&&tg(2\theta)(\la w_{0,0}|w_{1,1}\ra)+ctg(2\theta)\la w_{0,0}|w_{0,1}\ra 
+\\\nonumber&&
\la w_{0,0}|W_{0,1}\ra-
\la W_{0,0}|w_{0,1}\ra)-\\\nonumber
&&\la w_{0,0}|W_{1,1}\ra-
\la W_{0,0}|w_{1,1}\ra\\\nonumber
\end{eqnarray}}}

Thus, our bit escrow protocol gives quadratic sealing. 

\begin{remark}\label{remark:b}
Protocol \ref{pro:escrow} is sealing even if we modify 
it a little bit, as follows: 
at {\em revealing} time Alice first reveals $b$
and then Bob returns the qubit $q$.
In other words, if Bob has learned $\epsilon$ information
about $b$ after the deposit stage,  
then even if later on he gets to know $b$, he cannot avoid being 
detected with probability $\Omega(\epsilon^2)$. 
To see this, we use linearity. If Bob has a strategy which gives him 
detection probability $p$ in the modified protocol, 
then w.l.o.g. his strategy is to apply the identity if $b=0$   
and some unitary operation $U$ if $b=1$. However, since the 
$b=1,x=0$ and $b=1,x=1$ cases are linear combinations of the 
$b=0,x=0$ and $b=0,x=1$ cases, 
one can show that if Bob's probability for detection 
is $p$ in the $b=0$ case, then it is also $O(p)$ in the $b=1$ case, 
and therefore Bob does not have to apply $U$ in the first place. 
This means that if he has a cheating strategy for the modified protocol,
then he also has a cheating strategy with about the same parameters 
for protocol \ref{pro:escrow}, and so by Theorem \ref{thm:sealing}
the modified protocol is also quadratically secure. 
\end{remark}

\begin{remark}
One might suspect that this quadratic gap will always be the case 
for any reasonable set of vectors for Alice. 
This is not correct. If Alice only uses 
$\phi_{0,1}$ and $\phi_{1,0}$, then Bob has a strategy 
which gives him $^4\sqrt{p}$ advantage. 
We will not elaborate on this in this paper.
\end{remark}

\subsection{A Quadratic Strategy for Bob}

\begin{theorem}
\label{thm:quadraticbob}
Let $\rho_0, \rho_1$ be two density matrices of the same 
dimension, such that 
$\|\rho_0-\rho_1\|_t=t$.
Consider the following protocol. 
Alice tosses a random bit $b$. 
She chooses a pure state from the mixture 
 $\rho_b$, and sends it to Bob.
Then Bob returns to Alice the state,
and Alice projects it on the original state to test whether 
Bob has manipulated it. 
We claim that for any $1\ge p\ge 0$, there is a strategy for Bob 
such that he  learns $b$ with advantage 
$t\sqrt{p}$, and his probability of detection is at most
$\frac{1}{2}(1-\sqrt{1-p})$, which is $\Theta(p)$ for small $p$.
\end{theorem}

{\bf proof}:
Alice prepares an encoding $\psi_b$ of $b \in \{0,1\}$ in register
$B$, and sends register $B$ to Bob.
Let $\rho_b$ be the reduced density matrix of $\psi_b$ to register $B$.
We denote $t=\trn{\rho_0-\rho_1}$.
By Theorem \ref{thm:akn} we know that if Bob is interested in 
learning information about $b$, and is not concerned with
being detected cheating, the best he can do is a measurement
according to the eigenvalue basis of $\rho_0-\rho_1$. 
Given, any $0 \le p \le 1$ we modify this strategy to 
a strategy where 
the detection probability is at most $p$, 
and yet, Bob gets much information.

Let us consider more precisely 
Bob's best strategy for learning $b$ 
if he is not concerned with being caught. 
Let $\set{e_1,\ldots,e_K}$ be the eigenvector basis of $\rho_0-\rho_1$.
Let $V^+$ ($V^{-}$) be the set of eigenvectors $e$ with non-negative  
(negative) eigenvalues.
The measurement $M$ is defined by the Hermitian matrix for which 
$V^+$ is an eigenspace of eigenvalue $0$ and $V^-$
is an eigenspace of eigenvalue $1$. 
%This corresponds to the following decision procedure, in which 
%Bob applies when Bob gets a classical outcome $b\in \{0,1\}$ he guesses 
%that the density matrix is $\rho_b$. 
By Theorem \ref{thm:akn}
\begin{eqnarray}
\label{eq:trace}
|\trace{\rho_0M}-\trace{\rho_1M}| & = & {t \over 2}
\end{eqnarray}

To apply a weak form of the measurement $M$, 
Bob takes a one qubit ancilla $C$. 
He applies a unitary transformation $U$ on the received message and the 
ancilla, as follows:
\begin{eqnarray*}
U \ket{e,0} & = & 
\left\{
\begin{array}{ll}
\ket{e,0} & \mbox{If $e \in V^{+}$} \\
\ket{e} \tensor \ket{v} &
     \mbox{If $e \in V^{-}$}
\end{array}
\right.
\end{eqnarray*}
where $\ket{v}=\sqrt{1-p} \ket{0} + \sqrt{p} \ket{1}$ and $U$ is completed 
to a unitary transformation.
After applying $U$ Bob returns register $B$ to Alice, and keeps
the ancilla $C$ for himself.
Notice that the special case $p=1$ is equivalent to the measurement $M$. 

\begin{lemma}
$\trn{ U \rho_0 |_C - U \rho_1|_C } = t \sqrt{p}$.
\end{lemma}

\begin{proof}
We will show
\begin{claim}
\label{cl:urho0}
\begin{eqnarray*}
U\rho_0|_C & = & \trace{\rho_0M}|0\ra\la 0|+(1-\trace{\rho_0M})|v\ra\la v|) \\
U\rho_1|_C & = & \trace{\rho_1M}|0\ra\la 0|+(1-\trace{\rho_1M})|v\ra\la v|)
\end{eqnarray*}
\end{claim}

Thus, 
$U\rho_0|_C - U\rho_1|_C ~=~ 
(\trace{\rho_0M}-\trace{\rho_1M})(|0\ra\la 0|-|v\ra\la v|)
~=~ \pm \frac{t}{2}(|0\ra\la 0|-|v\ra\la v|)$,
where the last equality is due to Equation \ref{eq:trace}.
Since, $\trn{|0\ra\la 0|-|v\ra\la v|} =
2 \sqrt{1 -\la 0|v\ra|^2} = 2\sqrt{p}$ we get
$\trn{U\rho_0|_C - U\rho_1|_C} = t \sqrt{p}$ as desired.
\end{proof}

We now prove Claim \ref{cl:urho0}.

\begin{proof} (of Claim \ref{cl:urho0}).
We express
$\rho_0  =  \sum_j w_j|\alpha_j\ra\la \alpha_j| $,
where $\alpha_j$ is a pure state.We further express each $\alpha_j$
in the eigenbasis $\set{e_i}$: 
\begin{eqnarray*}
\ket{\alpha_j} &=& 
\sum_{i+} a^+_{ij} \ket{e_i^+}+
\sum_{i-} a^-_{ij} \ket{e_i^-}
\end{eqnarray*}
Applying $U$, this state is taken to:
 \begin{eqnarray*}
U \ket{\alpha_j,0} &=&
 \sum_{i+} a^+_{ij} \ket{e_i^+} \ket{0} 
+\sum_{i-} a^-_{ij} \ket{e_i^-} \ket{v}
\end{eqnarray*}
The reduced density matrix to the register $C$,
in case of event $\ket{\alpha_j}$ is:
\begin{eqnarray*}
& \sum_{i+} |a^+_{ij}|^2 |0\ra\la 0|+
\sum_{i-} |a^-_{ij}|^2 |v\ra\la v| &
\end{eqnarray*}
and altogether,
$U\rho_0|_C = 
\sum_j w_j (\sum_{i+} |a^+_{ij}|^2) |0\ra\la 0|+$ 

$\sum_j w_j (\sum_{i-} |a^-_{ij}|^2) |v\ra\la v|$.
To complete the proof we just notice that
$\sum_j w_j (\sum_{i+} |a^+_{ij}|^2) = \trace{\rho_0M}$.
The proof for $U\rho_0|_C$ is similar.
\end{proof}

We now analyze the error detection probability.

\begin{lemma}
$Prob(err) \le {1 \over 2}(1-\sqrt{1-p})$
\end{lemma}

\begin{proof}
Say Alice sent Bob the state $|w\ra$. We can express 
it as $\ket{w} = a \ket{w^+}+b \ket{w^-}$ where
$\ket{w^+} \in Span(V^+)$ and  $\ket{w^-} \in Span(V^-)$.
Bob applies $U$ on $w$ and gets
\begin{eqnarray*} 
U\ket{w} & = & a \ket{w^+,0} +b \ket{w^-,v} \\
& = & a \ket{w^+,0} +\sqrt{1-p} b \ket{w^-,0} + \sqrt{p} b \ket{w^-,1}
\end{eqnarray*}
Therefore, if we measure the last qubit, then with
probability $pb^2$ we end up in $\ket{w^{-}}$
and with probability $1-pb^2$ we end up in 
$a\ket{w^+} + \sqrt{1-p}b\ket{w^-}$
normalized.
Thus the density matrix of $U\ket{w}$
after tracing out the last qubit is:
\begin{eqnarray*}
\rho & = &
\left(\begin{array}{cc}
|a|^2 &
b\bar{a}\sqrt{1-p} \\
\bar{b}a\sqrt{1-p} &
|b|^2 
\end{array}
\right)
\end{eqnarray*}
To find out the probability for Alice not to detect Bob cheating, 
we calculate $\la w|\rho| w\ra$. 
We get:
\begin{eqnarray*}
Pr(\neg Err) & = & |a|^4+2|ab|^2\sqrt{1-p}+|b|^4\\&=&1-2|ab|^2(1-\sqrt{1-p})  
\end{eqnarray*}
The probability of Alice detecting an error 
is thus
$2|ab|^2(1-\sqrt{1-p}) \le \frac{1}{2}(1-\sqrt{1-p})$.
\end{proof}

\begin{remark}
The average of $|ab|$ can tend to  $0.5$, even when 
$t$ tends to $0$. This can be seen by taking $\rho_0$ to be composed 
of two states which are the basis states $\ket{0}$ and $\ket{1}$
rotated by $\theta$ towards each other, whereas $\rho_1$ is the mixture
of the basis states rotated by $\theta$ outwards. 
As $\theta$ tends to $0$, $t$ tends to $0$, but 
$|ab|$ tend to $0.5$. 
\end{remark}

\section{Proof of Theorem \ref{thm:bias}}

%\begin{proof} (Of Theorem \ref{thm:bias}: Biased coin flipping).
We show that no cheater can control the game.

\begin{description}
\item[When Bob cheats]:

Suppose Alice is honest and Bob is arbitrary.
Let us look at the mixture that Alice generates
at the first step of Protocol \ref{pro:bias}.
Let $\rho_{b=0}$ be the density matrix in the case $b=0$,
and $\rho_{b=1}$ in the case $b=1$.
Then
$\trn{\rho_{b=0}-\rho_{b=1}} = 2\cos(2\theta)$. 
It follows from Theorem \ref{thm:akn} 
that whatever Bob does, the probability that
$b'=b$ and Bob wins is at most
$\Pr(b'=b) \le {1 \over 2}+{\cos(2 \theta) \over 2} = \cos^2(\theta)$
which for $\theta ={\pi \over 8}$
is at most $0.86$.

\item[When Alice cheats]:

Now, suppose Bob is honest and Alice is arbitrary.
$\prob( \mbox{Alice wins })=x$, which is at most ${p_0+q_1 \over 2}$, 
whereas the probability that Alice loses is 
at least ${p_1 + q_0 \over 2}$. 
The difference  $|x-(1-x)|$ is at most
${p_0 - q_0 + q_1 - p_1 \over 2} \le {|p_0-q_0|+|p_1-q_1|\over 2} = 
|p_0-q_0|$, i.e., $x \le {1+|p_0-q_0| \over 2}$. 

Also, 
${p_{err}+q_{err}\over 2} \le 1-x$, as
whenever Alice is caught cheating
she loses. 
This implies that 
$\sqrt{p_{err}}+\sqrt{q_{err}} \le 2\sqrt{1-x}$
as the maximum is obtained when $p_{err}=q_{err}=1-x$.

Finally, from the proof of Theorem \ref{thm:binding} 
we have
$|p_0-q_0|\le { \sqrt{p_{err}} + \sqrt{q_{err}} \over cos(2\theta) }$.
Putting it all together we get:
\begin{eqnarray*}
x & \le & {1+|p_0-q_0| \over 2} \\
  & \le & {1 \over 2}+ 
          { \sqrt{p_{err}} + \sqrt{q_{err}} \over 2cos(2\theta) } \\
  & \le & {1 \over 2}+ {\sqrt{1-x} \over \cos(2\theta)}
\end{eqnarray*}
For $\theta={\pi \over 8}$ we get the quadratic
equation $4x^2+4x-7 \le 0$. Solving it we get 
$x \le {\sqrt{8}-1 \over 2} \le 0.9143$.
\end{description}

\bibliographystyle{plain}
%\bibliography{bibl}  

% ACM needs 'a single self-contained file'!
%\balancecolumns
%\appendix
%Appendix A
%\section{Headings in Appendices}

\end{document}